\documentclass[%
 reprint,
 amsmath,amssymb,
 aps, nofootinbib
]{revtex4-1}

\usepackage{graphicx}
\usepackage{bm}
\usepackage[dvipsnames, usenames]{xcolor}
\usepackage[normalem]{ulem}
\usepackage{booktabs}
\usepackage{multirow}

\begin{document}

\title{Radiation from Global Topological Strings using Adaptive Mesh Refinement: \\ Massive Modes}

\author{Amelia Drew}
 \email{a.drew@damtp.cam.ac.uk}
\author{E.P.S. Shellard}%
 \email{epss@damtp.cam.ac.uk}
\affiliation{%
 Centre for Theoretical Cosmology, Department of Applied Mathematics and Theoretical Physics,
University of Cambridge, Wilberforce Road, Cambridge CB3 0WA, United Kingdom
}%

\date{\today}

\begin{abstract}
We implement adaptive mesh refinement (AMR) simulations of global topological strings using the public code, GRChombo. We perform a quantitative investigation of massive radiation from single sinusoidally displaced string configurations, studying a range of string widths defined by the coupling parameter $\lambda$ over two orders of magnitude, effectively varying the mass of radiated particles $m_H \sim \sqrt{\lambda}$. We perform an in-depth investigation into the effects of AMR on massive radiation emission, including radiation trapping and the refinement required to resolve high frequency modes. We use quantitative diagnostic tools to determine the eigenmode decomposition, showing a complex superposition of high frequency propagating modes  with different phase and group velocities. We conclude that massive radiation is generally strongly suppressed relative to the preferred massless channel, with suppression increasing at lower amplitudes and higher $\lambda$. Only in extreme nonlinear regimes (e.g.\ with relative amplitude $\varepsilon \sim 1.5$ and $\lambda < 1$) do we observe massive and massless radiation to be emitted at comparable magnitude. We find that massive radiation is emitted in distinct high harmonics of the fundamental frequency of the string, and we  demonstrate that, for the sinusoidal configurations studied, massive radiation is exponentially suppressed with $\sqrt{\lambda}$ (i.e.\ the particle mass). Finally, we place these results in the context of axions and gravitational waves produced by cosmological cosmic string networks, and note that AMR provides a significant opportunity to explore higher $\lambda$ (thin string) regimes whilst using fewer computational resources.

\end{abstract}

\maketitle

\section{Introduction}

Topological or `cosmic' strings are predicted by many physically motivated field theories \cite{Kibble1976}, including grand-unified models and superstring theory. Usually arising as a result of a symmetry-breaking phase transition, they can lead to a wide variety of cosmological consequences \cite{Vilenkin:2000jqa}.  So-called `global' strings, which have a long-range Goldstone boson or axion field, are created from the breaking of a $U(1)$-symmetry with a single complex scalar field $\varphi$.  A key physical motivation for this scenario comes from the Peccei-Quinn $U_{\rm PQ}(1)$ symmetry, introduced to solve the strong CP problem of QCD \cite{Peccei1977a}. When $U_{\rm PQ}(1)$ is broken, axion strings are formed and become a potential source of dark matter axions \cite{Davis1986}. Both global and `local' gauged strings are a potential source of gravitational waves, with the potential for detection by LIGO-Virgo-KAGRA \cite{Abbott_2021}, LISA \cite{Auclair_2020, LISA}, NANOGrav \cite{NANOGRAV} or future gravitational wave experiments. Detection of cosmic strings would allow us to probe the symmetry breaking scale of the underlying high energy physics model.
 
Global cosmic strings have three potential radiative decay channels: massless axion (Goldstone) radiation, massive particle radiation and gravitational radiation. The balance between the massive and massless channels is determined partially by the symmetry-breaking potential $V(\varphi)$. In this work, we use $V(\varphi) = \frac{\lambda}{4}(\bar{\varphi}\varphi - \eta^2)^2$, where $\lambda$ is a positive constant and we set $\eta=1$. In numerical simulations, it is necessary to accurately resolve the small-scale dynamics of the string core in order to determine the relative significance of each of these channels. This is particularly important for massive particle radiation, as we expect this decay channel to be suppressed as the width of the string $\delta \sim 1/\sqrt{\lambda}$ decreases and the massive particle mass $m_{\mathrm H} \sim \sqrt{\lambda}$ increases. The spectrum of the propagating massive radiation will affect the relative magnitude of the massless (and gravitational) radiation channels for global strings, with knock-on effects e.g.\ for predictions of the QCD axion mass. The same issue arises in simulations of local cosmic strings but without the Goldstone boson decay channel, where any massive particle radiation will affect predictions of the gravitational wave spectrum.

Historically, cosmic string evolution has been modelled using the Nambu-Goto model (or the Kalb-Ramond model for global strings), which approximates strings as having an infinitely thin width. By construction, this approach does not model the massive decay mechanism as it integrates out internal degrees of freedom of the string. Massive particles are assumed to be too heavy to radiate, and are often quoted as being exponentially suppressed, for example in \cite{Olum2000, Moore:1998}. In contrast, it has been argued in other work, primarily from field theory simulations, that the massive decay channel may have a power law spectrum \cite{Vincent:1997cx}, which could be significant for non-linear string configurations. This has been a source of significant debate for both the estimation of the axion mass emitted from axion string networks \cite{Gorghetto_2021, Hindmarsh_2021, Buschmann_2022, Saurabh_2020, Hindmarsh2021} and the prediction of gravitational wave signatures from local cosmic strings \cite{Olum2000, Blanco_Pillado_2018, Matsunami_2019, Auclair_2020a, Hindmarsh_2020, Auclair_2021, Hindmarsh_2021a, Abbott_2021, Hindmarsh_2022}.

Resolution of realistic cosmic string widths in field theory simulations poses a very significant computational challenge. The ratio between the string width $\delta$ and the Hubble radius $\Lambda \lesssim H^{-1}$ is characterised by $\ln{\Lambda/\delta \sim 70}$ and $\ln{\Lambda/\delta \sim 100}$ for QCD axion and GUT scale strings respectively. However, typical field theory simulations using a fixed grid can only reach $\ln{R/\delta} \sim 8$, and must often employ numerical `tricks' in order to resolve the string core accurately as the background expands. The lack of dynamic range afforded by fixed grid simulations means that it is especially challenging to determine the massive radiation spectrum for realistic $\delta$. The lack of consensus on whether field theory simulations with low $\lambda$ can be reliably extrapolated to cosmological scenarios further complicates the above discussion. Adaptive mesh refinement (AMR) is a computational method that may allow us to probe a higher dynamic range using fewer computational resources, potentially providing the ability to make more concrete measurements of the $\lambda$-dependence of the massive radiation for string configurations of closer relevance for cosmological scenarios. AMR simulations of cosmic string loop collapse in full numerical relativity have been performed by other authors \cite{Helfer2019, Aurrekoetxea_2020, Aurrekoetxea_2022}.

%The primary analytic model has been Nambu-Goto evolution, for which the string core is approximated as having infinitely thin width. 

%The difference in scale between the typical string width $\delta$ and the Hubble radius $\Lambda \lesssim H^{-1}$ can create significant barriers to performing accurate simulations. as well as the large-scale evolution of the string network and propagation of emitted radiation. 

%As in \cite{Drew2019}, simulations are carried out using GRChombo with a coarse simulation box size of $256\times256\times32$ ($N_1 \times N_2 \times N_3$) and the same boundary conditions, resolution and regridding parameters; periodic boundary conditions in the $z$-direction and Sommerfeld boundary conditions in the $x$- and $y$- directions. In addition, approximately forty further simulations of global strings with finely spaced $\lambda$ values from $0.3 \le \lambda \le 2.8$ are presented, with a typical spacing $\Delta\lambda \approx 0.1$ to allow comparison with the analytically determined mass thresholds. These primarily use a coarse simulation box size of $256\times256\times32$, but we also investigate configurations with $256\times256\times16$ in order to facilitate larger relative amplitudes with more highly relativistic string configurations. \ad{General comment - is this reworded enough?} 

In this paper, we present a detailed analysis of the massive radiation from adaptive mesh refinement simulations of global cosmic strings. Section \ref{theory} outlines the theory of global string formation and Section \ref{numerics} gives details of the numerical implementation. Section \ref{convergencetesting} presents several convergence tests and detailed analysis of the reliability of AMR as a method for simulating massive radiation from global strings. Section \ref{massive} presents analytic models and numerical results for the relative amplitude of massive radiation when compared to massless radiation, as well as its power spectrum and $\lambda$-dependence. We conclude and discuss the implications of this work in Section \ref{conclusion}. We use `natural' units throughout, setting $\hbar = c = k_B = 1$ such that $[E] = [M] = [L]^{-1} = [T]^{-1}$.

%noting in particular the significant increase in complexity over the corresponding massless radiation

\section{Global String Theory and Radiation}\label{theory}

In this section, we provide a brief outline of the model for global cosmic strings and the radiation diagnostics used in this paper. Further information can be found in \cite{Vilenkin:2000jqa} and \cite{Drew2019}. 

The Goldstone model has a Lagrangian density $\mathcal{L}$ given by
\begin{equation}
    \mathcal{L} = (\partial_\mu\bar{\varphi})(\partial^\mu\varphi) - V(\varphi)\,,
    \label{GoldstoneLagrangian}
\end{equation}
with the potential
\begin{equation}\label{potential}
V(\varphi) = \frac{1}{4}\lambda(\bar{\varphi}\varphi - \eta^2)^2\,.
\end{equation}
The constant $\eta$ sets the symmetry breaking scale and, together with $\lambda$, the mass of the Higgs particle in the broken symmetry state, i.e. $m_{\rm H} = \sqrt{\lambda} \,\eta$, which emerges alongside the massless Goldstone boson. If we decompose the complex scalar field $\varphi$ into real and imaginary parts
\begin{equation}\label{complex}
\varphi = \phi_1 + i\phi_2\,,
\end{equation}
the Euler-Lagrange equations for the numerical evolution are given by 
\begin{equation}
    \frac{\partial^2\phi_{1,2}}{\partial t^2} -
    \nabla^2\phi_{1,2} + \frac{\lambda}{2}\phi_{1,2}(|\varphi|^2 - \eta^2) = 0\,.
     \label{EL}
\end{equation}
There exist vortex solutions to these equations in two dimensions, which extend to line-like global string solutions in three dimensions. A static \textit{ansatz} solution to (\ref{EL}) is given by
\begin{equation}
    \varphi(r, \theta) =  \phi(r)e^{in_w\theta} \label{phi}\,,
\end{equation}
where $\phi = |\varphi|$ and $n_w$ is the topological winding number, which we set to $n_w=1$. Substituting into the static part of the  Euler-Lagrange equations (\ref{EL}), this yields an ordinary differential equation which can be solved numerically to find the radial cross-section $\phi(r)$ (see \cite{Drew2019}). This two-dimensional cross section describes a defect with higher energy than the surrounding vacuum which, when extended to three dimensions, is known as a `global' cosmic string.

As discussed in \cite{Drew2019}, an oscillating global string will emit both massless (Goldstone) and massive (Higgs) radiation, for which there are significantly different analytic expectations. In order to analyse these separate modes, we must not only separate these from each other, but also disentangle the radiative modes from the string self-fields. We can rewrite the complex scalar field $\varphi$, defined by \eqref{phi}, as
\begin{equation}\label{Argand1}
\varphi(x^\mu) =  \phi(x^\mu)\,e^{i\,\vartheta(x^\mu)}\,,
\end{equation}
where both the magnitude $\phi (x^\mu)=|\varphi(x^\mu)|$ and the phase $\vartheta(x^\mu)$ are real scalar fields.
From this, it can be shown \cite{Drew2019} that direct numerical diagnostics for the distinct massive and massless components of the string energy-momentum tensor $T_{\mu\nu}$ can be defined using the real momenta and spatial gradients:
\begin{alignat}{4}\label{massivediagnostic}
 &{\Pi}_\phi& &\equiv{}& &{}\dot{\phi}{}& &= \frac{\phi_1\dot\phi_1 + \phi_2\dot\phi_2}{\phi}, \nonumber \\ 
 &{\cal D}_i\phi& &\equiv{}& &{}\nabla_i\phi{}& &= \frac{\phi_1\nabla_i\phi_1 + \phi_2\nabla_i\phi_2}{\phi}, \\
&{\Pi}_\vartheta& &\equiv{}& &{}\phi\,\dot{\vartheta}{}& &= \frac{\phi_1\dot\phi_2 - \phi_2\dot\phi_1}{\phi}, \nonumber \\ 
&{\cal D}_i\vartheta& &\equiv{}& &{}\phi\,\nabla_i\vartheta{}& &= \frac{\phi_1\nabla_i\phi_2 - \phi_2\nabla_i\phi_1}{\phi}.\label{masslessdiagnostic}
\end{alignat}
Here, $\Pi$ denotes a `momentum-like' quantity composed using time derivatives, where subscript $\phi$ denotes the massive radiation contribution and subscript $\vartheta$ denotes the massless radiation contribution. Similarly, $\mathcal{D}_i\phi$ and $\mathcal{D}_i\vartheta$ represent the $i$th components of vectors composed using `spatial-gradient-like' quantities, where $\phi=\phi(x^\mu)$ and $\vartheta = \vartheta(x^\mu)$ as defined by \eqref{Argand1}. These can be used to express the energy density in terms of massive and massless components in the following form:
 \begin{eqnarray}\label{energymomdiag}
   T^{00} =  {\Pi}_\phi^2 + ( {\cal D} \phi)^2 + {\Pi}_\vartheta^2 + ( {\cal D} \vartheta)^2 +  {\textstyle\frac{\lambda}{4}} (\phi^2 -\eta^2)^2.
\end{eqnarray}
We can also explicitly split the momentum component $T^{0i}$ of the stress tensor into massive and massless components, given by
\begin{eqnarray}\label{momentumdiag}
   P_i\equiv T^{0i} =  2({\Pi}_\phi {\cal D}_i \phi + {\Pi}_\vartheta{\cal D}_i \vartheta) \,,
\end{eqnarray}
where the two terms represent the massive and massless radiation energy fluxes respectively. This is analagous to the Poynting vector which describes radiation energy flux in electromagnetism. Choosing an outgoing radial direction in our cylindrical geometry, we can integrate the two components of ${\bf P}\cdot \hat{\bf r}$ on a distant surface to determine the energy flow out of an enclosed volume for each mode. The massive component is given by
\begin{equation}\label{Pmassive}
P_{\mathrm{massive}} \propto \int{(\Pi_{\phi}\mathcal{D}\phi)\cdot\mathbf{\hat{r}}\,\mathrm{d}S \mathrm{d}t}\,,
\end{equation}
and the massless component by
\begin{equation}\label{Pmassless}
P_{\mathrm{massless}} \propto \int{(\Pi_{\vartheta}\mathcal{D}\vartheta)\cdot\mathbf{\hat{r}}\,\mathrm{d}S \mathrm{d}t}\,.
\end{equation}

Finally and importantly, we note that the approximate width of the string core defined by the profile \eqref{phi} is given by  
\begin{equation}
\delta \approx m_{\rm H}^{-1} \equiv  (\sqrt\lambda\,\eta)^{-1}\,,
\end{equation}
where $m_\mathrm{H}$ is the mass of the Higgs particle $\phi$. For simplicity, we shall set $\eta = 1$, and rescale the mass $m_{\rm H}$ and the string width using only the parameter $\lambda$.
Exploring radiation of a wide range of masses obtained by varying $\lambda$ will form the basis for the analysis in this paper.

\section{Numerical Implementation}\label{numerics}\label{simulationsetup}

The simulations in this paper are performed using the adaptive mesh code, GRChombo \cite{Andrade:2021}. By using AMR, we are able to save computational time and resources compared to equivalent fixed grid simulations by resolving the string core at a higher refinement than parts of the simulation box at large distances from the string. This is particularly important for thin strings with high $\lambda \gtrsim 10$, where running accurate simulations in an appropriate amount of time (less than approximately a week)  becomes unfeasible.

Initial conditions are obtained in the same way as in \cite{Drew2019}, using dissipative evolution of a sinusoidal initial configuration 
\begin{equation}\label{sinusoidalmodel}
    {\bf{X}}(z) = \left(A\sin{\Omega_z z}, 0, z\right)\,
\end{equation}
from an initial amplitude $A$ that is $50\%$ larger than the target amplitude $A_0$. Here, $\Omega_z=2\pi/L$ is the fundamental frequency at small amplitude and $L$ is the wavelength of the string, equivalent to the $z$-dimension of the box. Radiation from the string is extracted on a cylinder at $R=64$ which is accurate to fourth order. This is a different method to that used in the analysis in \cite{Drew2019}, but the same as used in the convergence tests in the same paper. The evolution scheme is fourth-order Runge-Kutta, with fourth-order spatial discretisation. Further specific details about the AMR implementation and wave extraction are discussed in \cite{Radia2021}.

Production simulations with AMR are carried out using a coarse simulation box size of $256\times256\times32$ or $256\times256\times16$ ($N_1 \times N_2 \times N_3$), with periodic boundary conditions in the $z$-direction and Sommerfeld (outgoing radiation) boundary conditions in the $x$- and $y$- directions. A base grid of resolution $\Delta x_0 = 1$ is used with a base timestep $\Delta t_0 = \Delta x_0 / 4$. Each refinement level reduces both $\Delta t$ and $\Delta x$ by a refinement ratio of 2.

It is necessary to define a regridding threshold to determine where the adaptive mesh will refine within the simulation box. We define our regridding criterion to be 
\begin{equation}
    \Delta x \sqrt{(\nabla\phi_1)^2 + (\nabla\phi_2)^2} > |\phi_{\rm threshold}|\,,
\end{equation}
where $|\phi_{\rm threshold}|$ is a custom threshold input by the user and $\Delta x$ is the grid spacing on a specific refinement level. We choose $|\phi_{\rm threshold}| = 0.25$, with no enforced maximum level unless otherwise specified (for example, for convergence tests).

Finally, an important factor to consider in our simulations when analysing high frequency massive radiation is the use of Kreiss-Oliger dissipation. This is a numerical technique that is used to damp high frequency modes that can be generated when using finite difference methods \cite{Kreiss:1973}. It is often added to numerical simulations to ensure stability, as non-physical, high frequency modes can cause simulations to `crash' at late times. In our case, we must ensure that any dissipation applied does not interfere with physical high frequency radiation emitted from the string. GRChombo implements Kreiss-Oliger dissipation by adding the following term to the right side of the evolution equations: 
\begin{align}
    \frac{\sigma}{64\Delta x}\left(F_{i-3} - 6F_{i-2} + 15F_{i-1} - 20F_i
    +15F_{i+1}
    \right. \nonumber \\ \left.
    -6F_{i+2} + F_{i+3}\right)\,,
\end{align}
where $F$ is the relevant evolution variable, $i$ is the index for the grid point and $\sigma$ is the damping parameter set by the user \cite{Radia2021}. The parameter $\sigma$ must satisfy
\begin{equation}
    0\leq \sigma \leq \frac{2}{\alpha_C} \label{eq:CFL}
\end{equation}
for the evolution to be stable, where $\alpha_C = \Delta t / \Delta x$ is the `Courant-Friedrichs-Lewy' factor on a given refinement level. As outlined above, we set $\alpha_C = 1/4$ in our simulations. 

%We also compare these AMR runs to boxes without mesh refinement, in order to isolate the effects of regridding on the massive radiation spectrum and measured energy loss.

\section{Convergence Testing and Fixed Grid Comparison}\label{convergencetesting}

In this section, we investigate the convergence of our simulations, and the effects of AMR on the massive radiation emitted by oscillating strings when compared to fixed grid simulations. We choose to investigate $\lambda=1$ and $\lambda=2$ strings with $A_0=4$, an amplitude which is in the mildly nonlinear regime. We choose these two $\lambda$ because, as we will discuss in later sections, the mass dependence of the energy emitted via massive radiation around $\lambda=1$ remains consistent with analytic predictions, but we observe a change from the expected behaviour for $\lambda \gtrsim 1.5$. It is therefore important to characterise the radiation and any unphysical numerical effects in both regimes. We also investigate convergence for $\lambda=10$, an example which is investigated in detail in this paper and well into the regime where AMR effects are found to be significant.

We know from previous work \cite{Drew2019} that, unlike massless radiation, massive radiation shows some  sensitivity to the detail of the implementation of AMR regridding. Although the underlying reason is not entirely clear, it could be due to the averaging scheme used in GRChombo to pass data from finer to coarser refinement levels.  This introduces some small numerical errors which, although negligible in magnitude, are sufficient to affect measurements of the highly suppressed massive radiation from the string.  We note that spatial averaging will introduce a small first order contribution, which may slightly degrade the convergence of the fourth-order Runge-Kutta scheme and spatial derivative stencils used for the evolution. The boundaries between refinement levels can also potentially be a further source of numerical inaccuracy through reflection or resonance effects. For this reason, we investigate convergence both for simulations that use AMR and those that use a fixed resolution grid. We use the grid configurations presented in Table \ref{convergence_params}.

\subsection{Kreiss-Oliger Dissipation}

It is important to consider the effects of Kreiss-Oliger damping in our simulations. Any numerical effects introduced by the mesh refinement are of the order of the grid spacing $\Delta x$, and will be damped by any Kreiss-Oliger scheme. However, in our physical setup, we expect to observe physical high frequency massive radiation, including some modes that approach the $\sim\,1/\Delta x$ frequency targeted by the damping. It will therefore become impossible at a certain frequency to distinguish noise introduced by the refinement algorithm from physical radiation emitted from the string. This becomes more problematic as $\Delta x$ increases, as the dissipation will be applied to increasingly lower frequencies. We must therefore take care when applying Kreiss-Oliger dissipation to ensure that it interferes as little as possible with the physical radiation.

For a fixed grid and a string with low $\lambda$, we expect a low proportion of the signal to be emitted in high frequency modes. Therefore, for a base resolution on which the string is properly resolved, we expect to measure approximately the same $P_{\rm massive}$ independent of the $\sigma$ coefficient. This is exactly what we observe for $\lambda = 1$, where we see $\sim 0.6\,\%$ difference between $\sigma=0$ and $\sigma=1$ for $\Delta x=0.25$, and for $\lambda = 2$, where we see a difference of $\sim 3\,\%$ for the same parameters. As $\Delta x$ increases, we observe in both cases higher frequency physical modes being damped away, with increasingly lower frequencies affected as $\Delta x$ increases.

For AMR simulations, this interpretation becomes more difficult. We know that the AMR algorithm will introduce numerical noise, hence it is more important than for the fixed grid case to use damping. However, for $\Delta x = 1$, the base grid resolution used for most of our AMR simulations, we observe that Kreiss-Oliger dissipation directly damps all but the lowest frequencies of propagating modes. We therefore must decide carefully whether applying dissipation is appropriate. For $\lambda \lesssim 1.5$, we observe from the radiation spectrum that dissipation does not mitigate significantly against numerical effects from the mesh refinement, nor have any noticeable affect on $P_{\rm massive}$. This will be discussed further in Section \ref{convergencetestsresults}. For this case, we therefore decide not to employ dissipation, setting $\sigma = 0$. For $\lambda \gtrsim 1.5$, numerical effects from the regridding begin to affect the radiation signal. In this case, it is appropriate to implement damping, and we set $\sigma = 1$. However, we note in practice that, even in these cases, the application of damping has a very minimal effect on the final result.

\begin{table}[t]
    \centering
    \caption{Grid parameters for the convergence tests for the massive radiation $P_{\mathrm{massive}}$. We perform tests both with a fixed grid and with AMR. For the fixed grid test (\textit{FG}), the grid dimension $L_{\mathrm {max}}$ remains constant and the base grid resolution $\Delta x_0$ is changed. For the AMR test ($\textit{AMR}$), the maximum refinement level $l_{\rm max}$ is changed and $\Delta x_0$ remains constant. The base grid box resolution is given by $N_1 \times N_2 \times N_3$, with $(l_{\max}+1)$ total refinement levels including the coarsest base level, and grid spacings on the finest level given by $\Delta x_{l_{\rm max}}$. The grid parameters for the corresponding damping stages are identical, except that $l_{\max} = 1$ for the AMR runs and $l_{\max} = 0$ for the fixed grid.}
    \begin{ruledtabular}
    \begin{tabular}{cccccc}
        Test & $N_1 \times N_2 \times N_3$ & $l_{\max}$ & $L_{\max}$ & $\Delta x_0$ & $\Delta x_{l_{\rm max}}$  \\
                 \hline
        \textit{FG} & $256\times 256\times 32$ & - & 256 & 1 & - \\
        & $512\times 512\times 64$ & - & 256 & 0.5 & - \\
         & $1024\times 1024\times 128$ & - & 256 & 0.25 & - \\
         & $1536\times 1536\times 192$ & - & 256 & 0.167 & - \\
         & $2048\times 2048\times 256$ & - & 256 & 0.125 & - \\
         & $4096\times 4096\times 512$ & - & 256 & 0.0625 & - \\
        \hline
        \textit{AMR} & $256\times 256\times 32$ & 0 & 256 & 1 & 1 \\
        & $256\times 256\times 32$ & 1 & 256 & 1 & 0.5 \\
        & $256\times 256\times 32$ & 2 & 256 & 1 & 0.25 \\
        & $256\times 256\times 32$ & 3 & 256 & 1 & 0.125 \\
         & $256\times 256\times 32$ & 4 & 256 & 1 & 0.0625  \\
         & $256\times 256\times 32$ & 5 & 256 & 1 & 0.03125 \\

    \end{tabular}
    \end{ruledtabular}
    \label{convergence_params}
\end{table}

\subsection{Convergence Tests}\label{convergencetestsresults}

\begin{figure}
    \centering
    \includegraphics[width=0.5\textwidth]{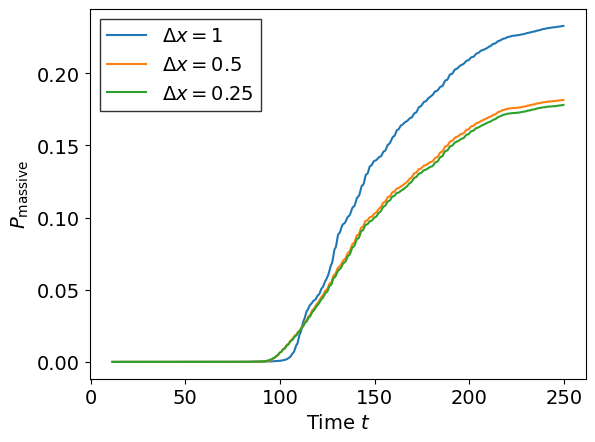}
    \includegraphics[width=0.5\textwidth]{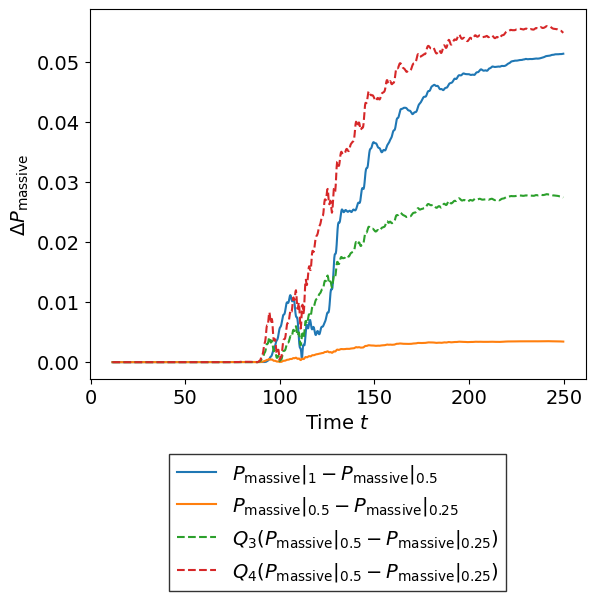}
    \caption{Absolute value (top) and convergence (bottom) of the energy emitted by massive radiation $P_{\mathrm{massive}}$ from a $\lambda=1$ string with initial amplitude $A_0=4$ and $\sigma=0$, measured on a cylinder at $R=64$ on a fixed grid for different refinements $\Delta x$ (test \textit{FG} in Table \ref{convergence_params}). The convergence plot shows the difference in the magnitude of $P_{\mathrm{massive}}$ between different resolutions, with the higher resolution results also plotted rescaled according to third- and fourth-order convergence.}
    \label{PoyntingFixedLambda1}
\end{figure}

Figure \ref{PoyntingFixedLambda1} shows the results of the convergence test for $\lambda=1$ using a fixed grid with the parameters in Table \ref{convergence_params} (test \textit{FG}) and $\sigma=0$. We test the cumulative massive component $P_{\mathrm{massive}}$ of the Poynting-like vector $\mathbf{P}$, determined using equation \eqref{Pmassive}. We observe that $P_{\rm massive}$ converges to a stable value by approximately $\Delta x = 0.25$, with approximately fourth order convergence. By fourth order Richardson extrapolation of the finest two simulations, we estimate the discretisation error at $t\sim 250$ to be $\Delta P_{\mathrm{massive}}/P_{\mathrm{massive}} \sim 0.1\,\%$.

\begin{figure}
    \centering
    \includegraphics[width=0.5\textwidth]{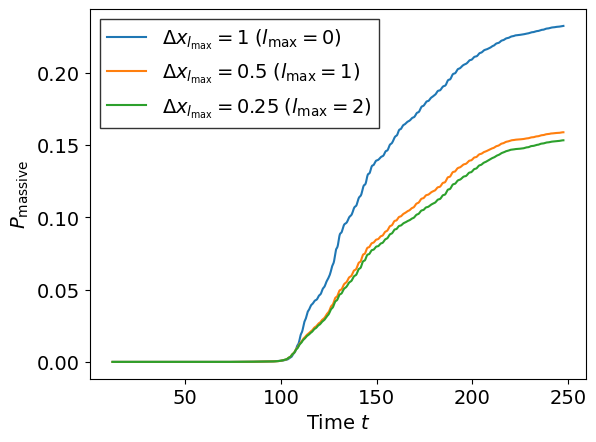}
    \includegraphics[width=0.5\textwidth]{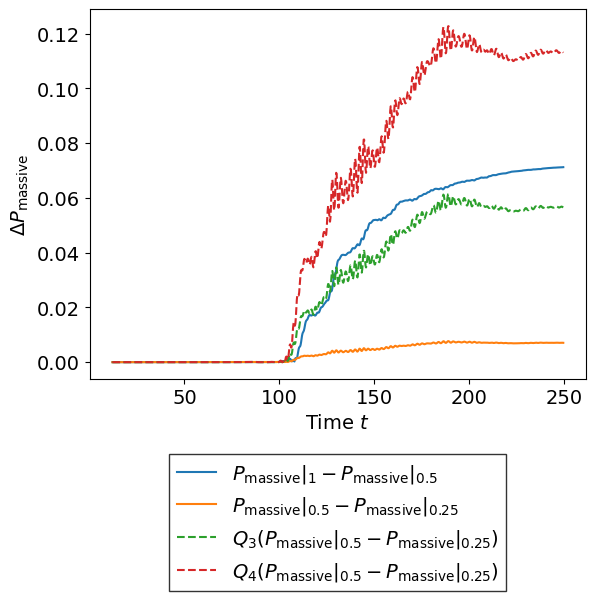}    
    \caption{Absolute value of the energy emitted by massive radiation $P_{\mathrm{massive}}$ from a $\lambda=1$ string with initial amplitude $A_0=4$, measured on a cylinder at $R=64$ using adaptive mesh refinement (test \textit{AMR} in Table \ref{convergence_params}). The convergence plot shows the difference in the magnitude of $P_{\mathrm{massive}}$ between different resolutions, with the higher resolution results also plotted rescaled according to third- and fourth-order convergence.}
    \label{PoyntingAMRLambda1}
\end{figure}

Figure \ref{PoyntingAMRLambda1} shows a convergence test for the same physical setup, but using mesh refinement. We set $|\phi_{\rm threshold}| = 0.25$, as used in the subsequent $\lambda=1$ simulations in this paper. We see again that $P_{\rm massive}$ converges to a stable value by $\Delta x_{l_{\rm max}} = 0.25$, where $\Delta x_{l_{\rm max}}$ is the grid spacing on the finest refinement level. This time, we observe approximately third-order convergence. As discussed at the start of the section, we note that this lower order is likely due to the mesh refinement averaging scheme beginning to affect the convergence. Although the frequency profile of the massive radiation is largely unaffected, there appears to be a small effective damping which reduces the overall magnitude of the convergent radiation amplitude by about $\sim 14\%$, relative to that from the fixed grid. Naively, we might assume that this is due to the coarser base grid being unable to resolve the high harmonics excited for massive radiation. However, the Nyquist frequency $F_N$ (highest frequency that can be recovered) for the base resolution $\Delta x_0 = 1$ is given by $F_N = 1/2\Delta x = 0.5$ in code units. As we will see in Section \ref{analyticexpectations}, the spatial `frequency' of the massive radiation for this configuration is given by $k_r/2\pi$, defined by equation \eqref{massivedispreln}. This allows harmonics of the fundamental frequency of the string up to $p\lesssim 16$ to be resolved, which is more than enough to accurately resolve the dominant propagating signal for $\lambda =1$. This therefore indicates that the reduction in $P_{\mathrm{massive}}$ measured is due to refinement level boundaries trapping some of the radiation that would otherwise propagate outwards.

For comparison, we run another convergence test for $\lambda=1$ with a lower regridding threshold $|\phi_{\rm threshold}| = 0.05$, which leads to larger areas being covered by each refinement level. We obtain results that are similar to $|\phi_{\rm threshold}| = 0.25$, but find that decreasing the regridding threshold increases the overall magnitude of the massive radiation, so that it is only $\sim 12\%$ lower than the fixed grid simulation. This demonstrates that increasing the size of the refinement areas around the string can ameliorate some of the effects of radiation `trapping', allowing more of the massive radiation to propagate outwards.

The equivalent convergence tests performed for $\lambda=2$ with $\sigma = 1$ demonstrate similar behaviour. For a fixed grid, we observe again that $P_{\rm massive}$ converges to a stable value by approximately $\Delta x = 0.125$, with approximately fourth order convergence. When using mesh refinement, $P_{\rm massive}$ again converges by $\Delta x_{l_{\rm max}} = 0.125$, this time with between second- and third-order convergence. Here, the overall magnitude of the energy is approximately equal to the fixed grid (see Figures \ref{PoyntingFixedLambda2} and \ref{PoyntingAMRLambda2} in Appendix \ref{AppendixA}).

Finally, Figures \ref{PoyntingFixedLambda10} and \ref{PoyntingAMRLambda10} in Appendix \ref{AppendixA} show a convergence test for $\lambda=10$ with $\sigma=1$. Figure \ref{PoyntingFixedLambda10} shows the fixed grid case, where $P_{\rm massive}$ converges to a stable value by approximately $\Delta x = 0.0625$, with approximately fifth order convergence. By fifth-order Richardson extrapolation, the discretisation error at $t\sim 200$ is approximately $\Delta P_{\mathrm{massive}}/P_{\mathrm{massive}} \sim 0.1\,\%$. Importantly, the final $P_{\mathrm{massive}}$ is $< 0.1\%$ of that emitted for $\lambda=1$, or $\mathcal{O}(10^3)\times$ smaller. This will be discussed further in Section \ref{relenergy}. Figure \ref{PoyntingAMRLambda10} shows a $\lambda=10$ convergence test with mesh refinement, again using $|\phi_{\rm threshold}| = 0.25$. $P_{\rm massive}$ again converges to a stable value by $\Delta x_{l_{\rm max}} = 0.0625$, and the overall magnitude of the radiation is $\sim 40\%$ lower than for the fixed grid, and with less than first-order convergence.\footnote{Note that the increase in radiation around $t\sim 200$ comes from radiation from the string, not the boundaries of the simulation.} We note that the convergence order is still increasing at the end of the simulation, so this may improve at later time. However, this demonstrates that, at high $\lambda$, the convergence is affected by numerical artefacts.

%\begin{figure}
%    \includegraphics[width=0.5\textwidth]{lambda_2_max_3_to_5_massive_convergence.png}
%    \includegraphics[width=0.5\textwidth]{lambda_2_max_3_to_5_massive_total.png}
%    \caption{Placeholder - absolute value of the $\{mn\} = \{2\,0\}$ Fourier mode of the massless radiation ${\cal{D}}\vartheta \cdot \hat{\bf r}$ from a $\lambda=10$ string with initial amplitude $A_0=4$, measured on a cylinder at $R=64$ for different maximum refinement levels $l_{\rm max}$ (test \textit{i)} in Table \ref{convergence_params}).}
%    \label{lambda10masslessmodesnoregridquadrupole}
%\end{figure}

%\begin{figure}
%    \includegraphics[width=0.5\textwidth]{lambda_2_full_refinement_massive_convergence.png}
%    \includegraphics[width=0.5\textwidth]{lambda_2_full_refinement_massive_total.png}
%\end{figure}

\section{Massive Radiation}\label{massive}

\subsection{Analytic Radiation Expectations}\label{analyticexpectations}

In this section, we determine the analytically predicted mode decomposition of massive radiation from global strings. We outline the properties of massive radiation, particularly the thresholds in $\lambda$ that determine whether certain modes are able to propagate and their dependence on the string amplitude. We also describe its complex wavepacket structure and outline a method to analytically separate propagating radiation from self-field modes.

\subsubsection{Massive Thresholds}

The presence of massive modes radiated from global strings can be demonstrated by linear expansion of the field equations (\ref{EL}) around the vacuum state $|\varphi|=\eta=1$. We first define the general form of the Argand representation 
\begin{equation} \label{Argand}
    \varphi(x^\mu) =  \phi(x^\mu)\,e^{i\,\vartheta(x^\mu)}\,,
\end{equation}
where both the magnitude $\phi (x^\mu)=|\varphi(x^\mu)|$ and the phase $\vartheta(x^\mu)$ are real scalar fields. Substituting into \eqref{EL}, the real part of the field equations is given by
\begin{equation}\label{real}
\frac{\partial^2\phi}{\partial t^2} -\nabla^2\phi = \phi \left [\left(\frac{\partial\vartheta}{\partial t}\right)^2-(\nabla\vartheta)^2+{\frac{\lambda}{2}} (1-\phi^2)\right]\,.
\end{equation} 
Assuming that $\vartheta$ is nearly constant far from any strings, (\ref{real}) becomes
\begin{equation}
    \frac{\partial^2\phi}{\partial t^2}-\nabla^2\phi - {\frac{\lambda}{2}}  \phi \left(1-\phi^2\right) = 0\,.  \label{massivediag}
\end{equation}
Expanding around the vacuum state $|\varphi|=\eta$ (where we have taken $\eta=1$) using $\phi = 1 + \chi$, it can be demonstrated that massive radiation obeys the Klein-Gordon equation
\begin{equation}\label{KleinGordon2}
    \frac{\partial^2\chi}{\partial t^2}-\nabla^2\chi + m_H^2\chi = 0\,,
\end{equation}
with $m_H = \sqrt\lambda\, \eta$. 

As discussed in \cite{Drew2019}, the massless radiation component can be decomposed into separable eigenmodes denoted by eigenvalues $\{pmn\}$, where $p,\,m$ and $n$ are positive integers used to denote the harmonics in $t$, $\theta$ and $z$ respectively. The radial wavenumber $\kappa_{pn}$ for each mode can be calculated as a function of the fractional increased path length $\alpha= T/L$, defined to be the path length of the string $T$ (which also determines its period of oscillation) relative to its periodicity $L$. The wavenumber can then be used to determine whether or not a certain mode of radiation will propagate. The radiation of massive modes from an oscillating global string is qualitatively different to massless radiation, due to the presence of the mass threshold $m_H =\sqrt\lambda\, \eta$. A sinusoidal string solution radiates into the lowest massless quadrupole mode $\{220\}$ for any initial amplitude. In contrast, massive modes must be sufficiently energetic to become propagating radiation with the lowest available mode depending on the mass threshold. This can be demonstrated similarly by deriving an expression for the massive radial wavenumber. Equation (\ref{KleinGordon2}) can be rewritten in cylindrical coordinates as with the massless case, obtaining
\begin{equation}\label{massivewaveeqn}
\frac{\partial ^2 \chi}{\partial t^2} - \frac{\partial ^2 \chi}{\partial r^2}- \frac{1}{r}\frac{\partial  \chi}{\partial r} - \frac{1}{r^2}\frac{\partial ^2 \chi}{\partial \theta^2} - \frac{\partial ^2 \chi}{\partial z^2} + m_H^2\chi  = 0\,.
\end{equation}
This is soluble using separable methods with the \textit{ansatz} $\chi(t,r,\varphi ,z) = T(t)\,R(r)\, \Theta (\theta )\, Z(z)$ to find asymptotic massive radiation modes. Substituting into (\ref{massivewaveeqn}), we obtain
\begin{equation}\label{separationmassive}
    \frac{T''(t)}{T(t)} -  \frac{R''(r)+ R'(r)/r}{R(r)}-\frac{1}{r^2 }\frac{\Theta ''(\theta )}{\Theta (\theta )} - \frac{Z''(z)}{Z(z)} + m_H^2=0\,.
\end{equation}
Rearranging and substituting appropriate separation constants, we obtain
\begin{equation}\label{radialdepmassive}
    \frac{R''(r)+ R'(r)/r}{R(r)}-\frac{m^2}{r^2 }=-\omega_p^2 + k_z^2 + m_H^2 = -k_r^2\,.
\end{equation}
where $\omega_p = 2\pi p/\alpha L = \Omega_z p/\alpha$ represents the $p^{\rm th}$ harmonic of the oscillating string, $k_z = \Omega_z n$ is the wavenumber in the $z$-direction and $k_r$ is the radial wavenumber. From this, we deduce that the massive modes obey the dispersion relation
\begin{equation}\label{dispersionrelationmassive}
    \omega_p^2 ~=~  k_r^2 + k_z^2 + m_H^2\,,
\end{equation}
which implies
\begin{equation}\label{massivedispreln}
 k_r \equiv \Omega_z\kappa_{pn} = \Omega_z \sqrt{\left({p}/{\alpha}\right)^2 - n^2 - m_H^2 /\Omega_z^2}\,.
\end{equation}
Radiation can radially propagate only if $k_r$ is real, so from (\ref{massivedispreln}), we obtain the expression for the lowest propagating harmonic
\begin{equation}\label{massthresholdlow}
p _{\rm min}\;> \;\alpha \sqrt {m_H^2/\Omega_z^2 + n^2} \; \approx\; m_H/\Omega_z\,,
\end{equation}
where in the last expression we have assumed that $L \gg m_H^{-1}$ and that $\alpha$ is close to unity.\footnote{We note that, in principle, the quadrupole $\{pmn\}=\{p_{\rm min}2\,0\}$ is the lowest massive harmonic available at a given order $p$. However, we shall see in practice that the dipole $\{p_{\rm min}\,1\,1\}$ is favoured when also above threshold.} As $\lambda$ increases and for a fixed $L$, a higher $p_{\rm min}$ is required to overcome the mass threshold and allow massive radiation to propagate. This effectively cuts off modes at lower frequencies, as they become evanescent. In order to determine the exact dependence of the massive spectrum on $\lambda$, equation \eqref{massivedispreln} can be rearranged as follows:
\begin{equation}
    \label{lambdathresholds}
    \lambda < \lambda_{pn} = \left(\frac{2\pi}{L}\right)^2\left(\frac{p^2}{\alpha^2} - n^2\right) \,,
\end{equation}
where $\lambda_{pn}$ is the threshold that $\lambda$ must (perhaps counter-intuitively) be below for a given mode $\{pn\}$ to propagate.

\subsubsection{Calculating $\alpha$}\label{alphacalc}

In order to calculate the values of $\lambda_{pn}$, it is necessary to calculate the fractional increase in path length $\alpha$ of the displaced string relative to the periodicity in the $z$-direction $L$. As outlined in the previous section, this is simply defined by $\alpha = T/L$ where $T$ is the path length, which also determines the period of oscillation of the string. To contextualise this calculation, we first recall from \cite{Battye1993, Battye1995} that the solution for a displaced string in the Nambu-Goto model is given by the expression
\begin{equation}\label{staticX}
{\bf X}(s, \varepsilon) = \bigg{(}\varepsilon\cos{s},0,\int_0^{s}{\sqrt{1-\varepsilon^2\sin^2{\theta}}\,\mathrm{d}\theta}\bigg{)}\,,
\end{equation}
where we have set $\Omega = 2\pi/T = 1$, the invariant amplitude $\varepsilon = 2\pi A/T$ with $0 \leq \varepsilon \leq 1$ and $0 \le s \le 2\pi$ is a parameter along the string over a single period. However, as we are evolving the full field equations, it is not a given that this analytic Nambu-Goto solution is `correct'. For this reason and for computational convenience, we use sinusoidal initial conditions which are damped to an appropriate intermediate configuration, which may or may not correspond to either a sinusoidal or Nambu-Goto model. We note in any case that a sinusoidal model is approximately equivalent to the Nambu-Goto model at low amplitudes $\varepsilon \ll 1$. However, it is useful to calculate $\alpha$ directly for both models, which we expect to provide upper and lower bounds for the damped solution.

We begin by calculating $\alpha$ for the simpler case of a sinusoidal string. The string displacement in the $x$-direction is given by 
\begin{equation}\label{sinusoidalmodelcopy}
{\bf{X}}(z) = \left(A\sin{\Omega_z z}, 0, z\right)\,.  
\end{equation}
We can determine the total path length $T$ for one period of the sinusoidal string from the simple integral
\begin{align}
    T_{\rm sin} &= \int^{L}_0{ \sqrt{1+\left(\frac{\partial x}{\partial z}\right)^2}\,\mathrm{d}z}\,,
%     &= \int^{L}_0{\sqrt{1+\Omega_z^2 A^2\cos^2{\Omega_z z}}\; \mathrm{d}z}\,,
%     &= \int^{2\pi}_0{\sqrt{1+\epsilon^2\cos^2{z}}\; dz}\,,
\end{align}
where we integrate from $0 \leq z \leq L$. Substituting the configuration (\ref{sinusoidalmodelcopy}), the increased path length $\alpha_{\rm sin}$ is then calculated simply as
\begin{equation}\label{sinusoidal}
    \alpha_{\rm sin} = \frac{1}{L}\int^{L}_0{\sqrt{1+\Omega_z^2A^2\cos^2{\Omega_z z}}\; \mathrm{d}z}\,.
\end{equation}
The path length $T_{\rm sin} = T_{\rm sin}(A)$ is a function of amplitude, with the $z$-periodicity being fixed at a constant $L$. We also note that, in this model, it is possible to create initial conditions that have an effective $\varepsilon_{\mathrm eff} > 1$, which we use to probe extreme nonlinear regimes in Section \ref{relenergy}.

\begin{figure}
    \centering
    \includegraphics[width=\linewidth]{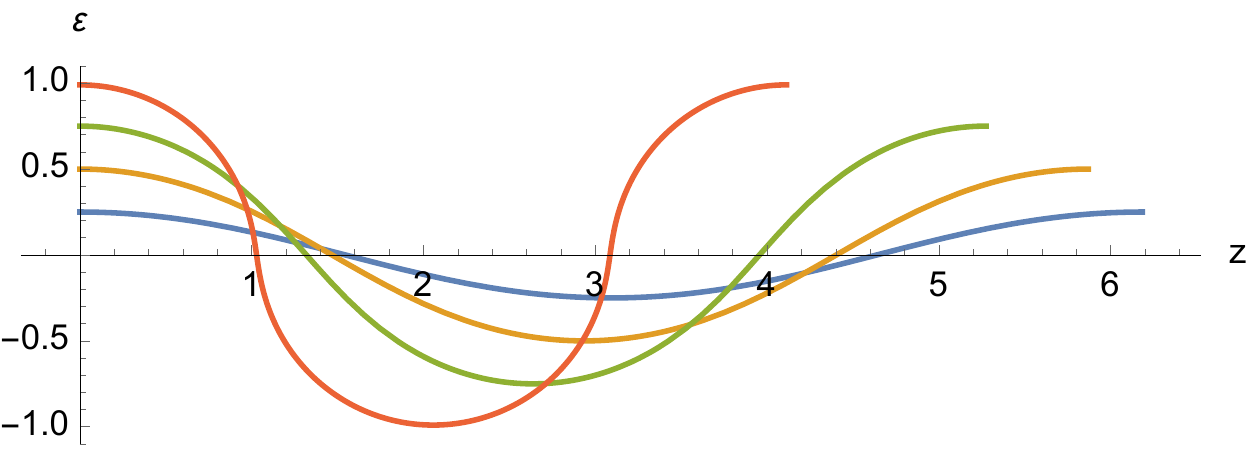}
    \caption{Parametric plot of accurate Nambu-Goto string initial conditions for amplitudes $\varepsilon=0.25$ (blue), $\varepsilon=0.5$ (yellow), $\varepsilon=0.75$ (green) and $\varepsilon=1$ (red).}
    \label{MathematicaGraph}
\end{figure}

In the Nambu-Goto case (\ref{staticX}), the calculation is less obvious due to the parametrisation by $s$. The path length is calculated using the integral
\begin{align}
    T_{\rm NG} &= \int^{2\pi}_0{ \sqrt{\left(\frac{\partial x}{\partial s}\right)^2+\left(\frac{\partial z}{\partial s}\right)^2}\; \mathrm{d}s} ~=~ 2\pi\,,
\end{align}
where we integrate over $0 \leq s \leq 2\pi$ for a single period. In this case, the string is fixed to be of parametric length $2\pi$, so when the amplitude of the string is increased, the periodicity in $z$ decreases accordingly. We therefore have the opposite situation to the sinusoidal case, which has a variable parametric path length that depends on the amplitude and fixed periodicity in $z$. Importantly, in the sinusoidal model, $A$ can be chosen to have any value without changing the $z$-periodicity $L$, whereas for Nambu-Goto strings, $L=z(2\pi, \varepsilon)$ is analytically determined by the model via the fixed path length. This is demonstrated by Figure \ref{MathematicaGraph}, which shows a parametric plot for four different invariant amplitudes $\varepsilon=0.25,\,0.5,\,0.75$ and $1.0$, demonstrating the decrease in $z(2\pi, \varepsilon)$ with increasing amplitude. From equation (\ref{staticX}), the periodicity $L$ in $z$ is given by 
\begin{equation}
    z(2\pi,\,\varepsilon) = \int_0^{2\pi}{\sqrt{1-\varepsilon^2\sin^2{\theta}}\,\mathrm{d}\theta}\,,
\end{equation}
where $0 \le \varepsilon \le 1$. The increase in path length $\alpha_{\rm NG}$ is given by the ratio of the path length $T_{\rm NG}$ to the periodicity in the $z$-direction as
\begin{align}\label{alpha}
     \alpha_{\rm NG} &= \frac{T_{\rm NG}}{z(2\pi,\varepsilon)} ~=~ \frac{2\pi}{\int_0^{2\pi}{\sqrt{1-\varepsilon^2\sin^2{\theta}}\,\mathrm{d}\theta}}\,.
\end{align}

To compare these two models in the context of our simulations, we need to compute $\alpha$ for different amplitudes $A$ and a fixed $z$-periodicity $L$. Fixing $L$ in the above Nambu-Goto model necessarily means that $\varepsilon$ is determined by the model for a given $A$. We define $A_{\,\rm{rel}}$, the amplitude relative to the $z$-periodicity, as
\begin{equation}\label{rescale}
    A_{\,\rm{rel}} = \frac{4A}{z(2\pi,\varepsilon)} = \frac{4A}{L}\,,
\end{equation}
so that in the limit $A=L/4$, we have $A_{\,\rm{rel}} = \varepsilon = 1$. This equation is implicit in $\varepsilon$ and must be solved numerically to find the desired $\varepsilon$ such that $z(2\pi,\varepsilon) = L$.

Comparing values of $\alpha$ for different $A_{\,\rm{rel}}$, we observe from Table \ref{alphavalues} that there is an additional path length contribution from the Nambu-Goto model compared to the sinusoidal approximation, which increases as $A_{\,\rm{rel}}$ increases. We will see in Section \ref{massiveanalysis} that this difference is significant when calculating the harmonic thresholds $\lambda_{pn}$ for the massive radiation.

\begin{table}
\caption{Fractional path length increase $\alpha$ for the Nambu-Goto ($\alpha_{\rm NG}$) and sinusoidal ($\alpha_{\rm sin}$) models for a range of relative amplitudes $A_{\,\rm{rel}}$.}
\centering
\label{alphavalues}
\begin{ruledtabular}
\begin{tabular}{l | c c}

$A_{\,\rm{rel}}$ & \multicolumn{2}{c}{$\alpha$ model} \\ 
& Sinusoidal $\alpha_{\rm sin}$ & Nambu-Goto $\alpha_{\rm NG}$ \\

\hline

0 & 1 & 1 \\

0.1 & 1.00614 & 1.00616 \\

0.25 & 1.0375 & 1.0382 \\

0.5 & 1.13984 & 1.14909 \\

0.75 & 1.28729 & 1.32541 \\

0.875 & 1.37264 & 1.43739 \\

0.95 & 1.42666 & 1.51362 \\

1.0 & 1.4637 & 1.5708 \\

\end{tabular}
\end{ruledtabular}
\end{table}

%\begin{table}
%\caption{Fractional path length increase $\alpha$ for the Nambu-Goto and sinusoidal models for a range of relative amplitudes $\epsilon$.}
%\centering
%\label{alphavalues}
%\begin{tabular}{l c c}
%\toprule
%$\epsilon$ & $\alpha$ \\ 
%\midrule
%0 & 1 \\

%0.1 & 1.00616 \\

%0.25 & 1.0382 \\

%0.5 & 1.14909 \\

%0.75 & 1.32541 \\

%0.875 & 1.43739 \\

%0.95 & 1.51362 \\

%1.0 & 1.5708 \\
%\bottomrule

%\end{tabular}
%\end{table}

\subsection{Radiation Properties}\label{radproperties}

Having derived an expression for the lowest propagating harmonic $p_{\rm min}$ for a given amplitude and $\lambda$ \eqref{massthresholdlow}, in this section we explore the properties of the different modes of massive radiation. Unlike massless radiation, it can be shown from the dispersion relation (\ref{dispersionrelationmassive}) that massive radiation has a separate phase velocity $v_{\rm ph}$ and group velocity $v_{\rm g}$, 
\begin{equation}
v_{\rm ph} = \frac{\omega}{k} \,,\qquad \qquad v_{\rm g} = \frac{d\omega}{dk}\,,
\end{equation}
with the latter representing the speed of energy transfer (focusing primarily on the radial component). The radial propagation velocity of the dominant massive modes is generically well below the speed of light, depending on how close the $p _{\rm min}$ harmonic is to the mass threshold. For example, for a string of unit mass ($\lambda=1$) and oscillation periodicity $L=32$ ($\Omega_z=0.2$), the lowest propagating harmonic is $p_{\rm min} =6$ with the quadrupole $\{6\, 2\,0\}$ having $v_{\rm g} =0.51$, and the dipole $\{6\, 1\,1\}$ about 5\% slower at $v_{\rm g}= 0.48$ (i.e. both at approximately half the speed of light).   In principle, lower massive harmonics $p<p_{\rm min}$ will oscillate as evanescent waves, representing   a `self-field' (bound modes) moving with the string but not propagating away. However, these asymptotic evanescent modes predicted by \eqref{massivedispreln} are exponentially suppressed over short lengthscales, so any massive self-field modes present at large radius are better understood as a response to the long-range massless self-field (see Section \ref{selffieldsection}).

Given that massive string radiation is typically a high harmonic of the driving frequency $\Omega_z$, we expect its generation mechanism to be highly nonlinear and dependent on self-interaction terms. The fundamental frequency for our sinusoidal string solution is generically well below the mass threshold $\Omega_z \lesssim m_H$ required for propagation, so any radiation modes will be strongly suppressed, given the high-order interactions required for their creation. For small oscillations $\varepsilon\ll 1$, we expect the radiation amplitude to be suppressed as an exponential of the radiating harmonic $p$ or, alternatively, the string curvature scale $R$ (see Section \ref{expdep}).\footnote{Here, $R$ is not to be confused with the extraction radius.} For this reason, we can anticipate that any massive radiation present will be dominated by the lowest time harmonic available $p_{\rm min}$.  

\subsubsection{Separation from Self-Field}\label{selffieldsection}

In addition to the propagating modes discussed above, we can identify the presence of massive self-field modes as a solution to the massive mode equation (\ref{real}). This requires us also to understand the form of the self-field of the massless radiation, $\vartheta_{\rm sf}$. 

\begin{figure}
    \centering
    \includegraphics[width=\linewidth]{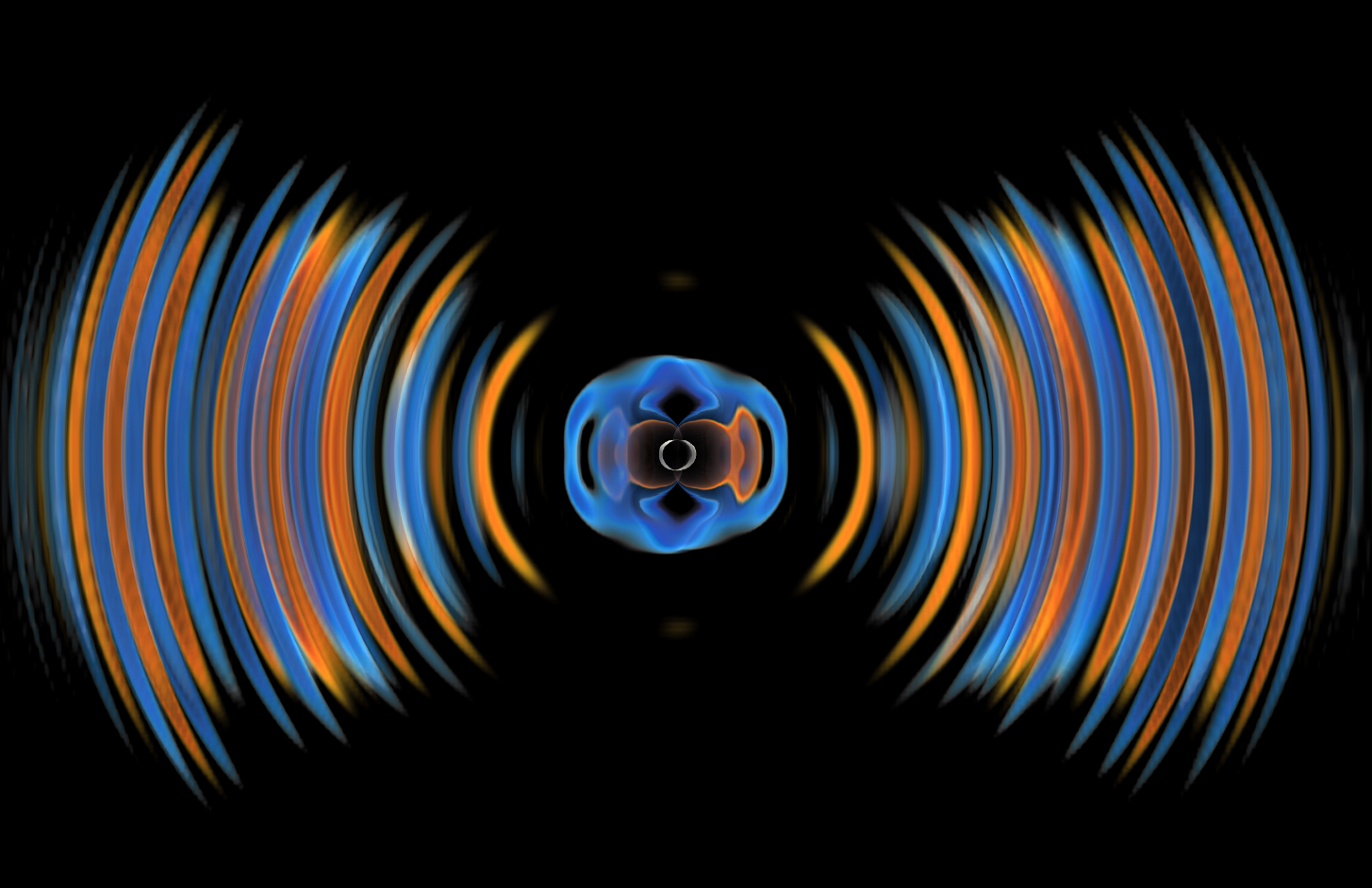}    
    \includegraphics[width=\linewidth]{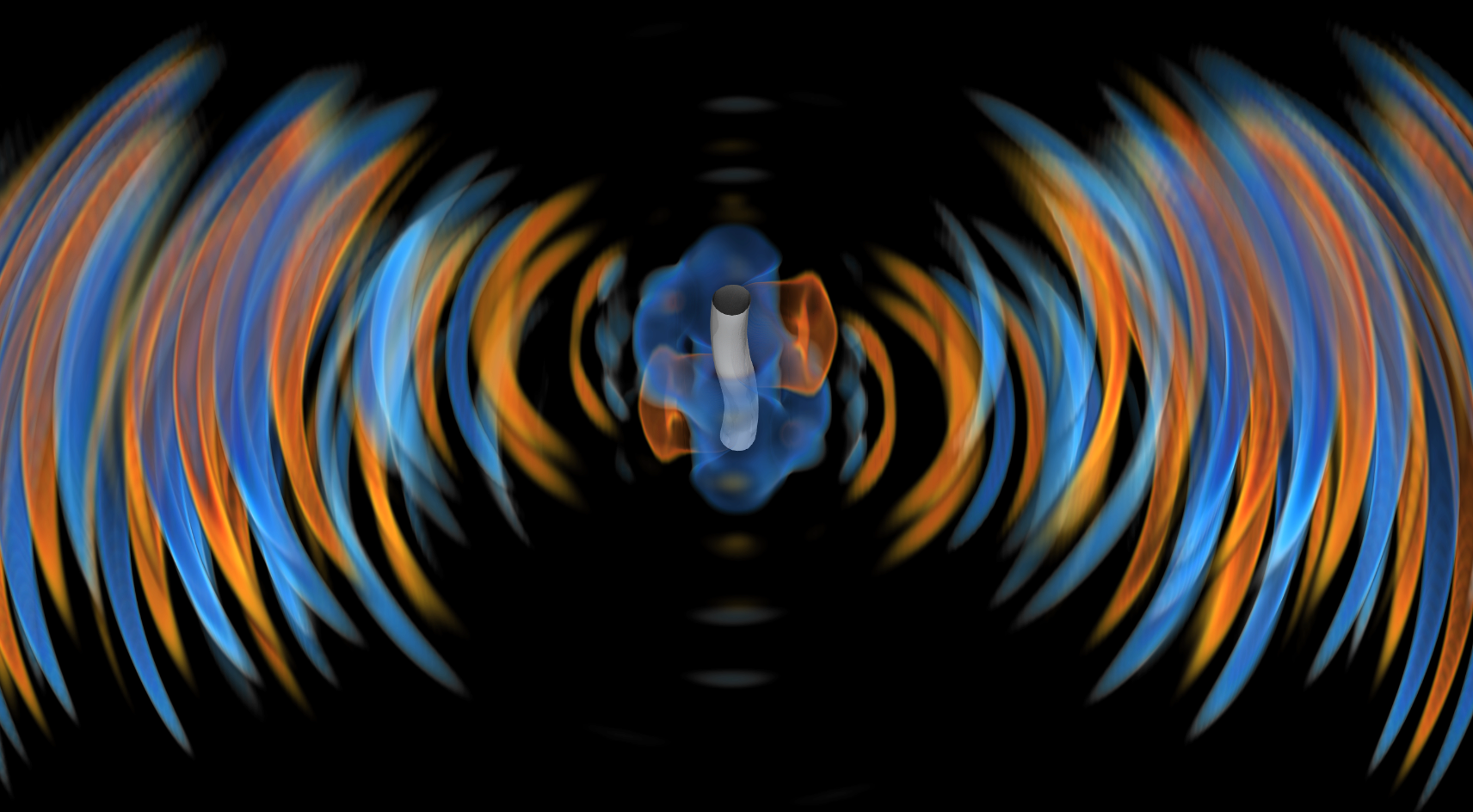}
    \caption{Volume rendering in 3D space $(x,y,z)$ of the massive radiation $\Pi_\phi$ from a $\lambda = 1$ string with initial amplitude $A_0 = 4$. The radiation is emitted from a string at the centre of the grid. The lowest propagating dipole eigenmode $\{pmn\} = \{6\, 1\, 1\}$ is dominant, but the different phase and group velocities give rise to a more complex structure of outgoing wavepackets.}
    \label{ParaviewMassiveScreenshotAmp4}
\end{figure}

As discussed in \cite{Drew2019}, at small amplitude ($\varepsilon \ll 1$), the sinusoidal string initial conditions (\ref{sinusoidalmodelcopy}) with the string field \textit{ansatz} (\ref{phi}) yields an approximate massless self-field $\vartheta_{\rm sf}(t, \bf{x})$ of the following form:
\begin{eqnarray}
\label{selffieldapprox}
\vartheta_{\rm sf}(t,{\bf x}) \approx \tan^{-1} \left ( y/ X(t,{\bf x}) \right)\,,\nonumber\\
X(t,{\bf x}) = x - A \cos\Omega_z t \, \sin\Omega_z z\,, 
\end{eqnarray}
valid in the region $A\ll r \lesssim {\cal O}(\hbox{few}\times L)$, where $X(t,{\bf x})$ is the $x$-coordinate relative to the string core. We can substitute derivatives of $\vartheta_{\rm sf}$ in the time-varying source term $\dot\vartheta^2 - (\nabla\vartheta)^2$ on the right hand side of \eqref{real}, copied again below for ease of reference:
\begin{equation}\label{realcopy}
\frac{\partial^2\phi}{\partial t^2} -\nabla^2\phi = \phi \left [\left(\frac{\partial\vartheta}{\partial t}\right)^2-(\nabla\vartheta)^2+{\frac{\lambda}{2}} (1-\phi^2)\right]\,.
\end{equation} When measured on a distant cylinder at fixed radius $R$, the self-field dipole from the time derivative term $\dot\vartheta_{\rm sf}^2$ is considerably larger than from the radial derivative $(\partial \vartheta_{\rm sf}/\partial r)$. However, we must include contributions from the angular derivative $\partial \vartheta_{\rm sf}/\partial \theta$ and from the $z$-direction $\partial \vartheta_{\rm sf}/\partial z$. The leading source contribution is the static term $((1/r)|\partial\varphi/\partial\theta|)^2 = (n_w\phi/r)^2$, arising from the angular derivative. This means that $\phi$ approaches the vacuum state with an asymptotic power law $\phi \sim 1 - r^{-2}$, rather than exponentially as would be expected for a massive field. The leading-order time-varying source contributions to the massive field equation (\ref{realcopy}) are then, using $\phi = 1+\chi$:
%\begin{eqnarray} \label{massiveselffield}
%\frac{\partial^2\chi}{\partial t^2} -\nabla^2\chi ~&=&~ - {\textstyle\frac{A^2\, \Omega^2}{4 r^2}} \left(1 - \cos 2 \theta\right ) \sin 2\Omega t\nonumber\\
%&&~+ {\textstyle\frac{2A}{r^3}} \cos \theta \, \sin \Omega z \, \sin \Omega t\,.
%\end{eqnarray}
\begin{eqnarray} \label{massiveselffield}
\frac{\partial^2\chi}{\partial t^2} -\nabla^2\chi ~&=&~ - {\textstyle\frac{A^2\, \Omega^2}{4 r^2}} \left(1 - \cos 2 \theta\right ) \cos 2\Omega_z t\nonumber\\
&&~+ {\textstyle\frac{2A}{r^3}} \cos \theta \, \sin \Omega_z z \, \cos \Omega_z t\,.
\end{eqnarray} The first source term arises directly from the square of the dipole term, so the time periodicity is that of the second harmonic, while $(\sin \theta)^2$ splits into  monopole and quadrupole contributions, with no $z$-dependence after adding $(\partial \vartheta/\partial z)^2$. The second line has a dipole cross term from $(\partial \vartheta/\partial \theta)^2$ which has the original time, angle and $z$-dependence of the string source.  Given the simplicity of the linearised wave equation (\ref{massiveselffield}), the solutions (and first derivatives, like $\Pi_\phi$ in (\ref{massivediagnostic})) will inherit the same $t$, $\theta$ and $z$-dependence as the right-hand side, whatever the resulting radial profile.  This means that in any FFT analysis we can expect a non-propagating massive self-field to be present, contributing to the monopole  $\{2\, 0\,0\}$, quadrupole $\{2\, 2\,0\}$ and dipole $\{1\, 1\,1\}$ eigenmodes. We also note that there are well-known radial oscillation modes in the string width which can, in principle, create a small monopole mode.

%\subsection{Energy Loss}

%It can be shown that radiation lost through massive radiation obeys an exponential dependence on $\lambda$. Exponential dependence of massive radiation on $\lambda$ \cite{Olum2000} - derive an exponential dependence on the mass (lambda in eqn. (7))

\begin{figure}[!]
    \centering
    \includegraphics[width=\linewidth]{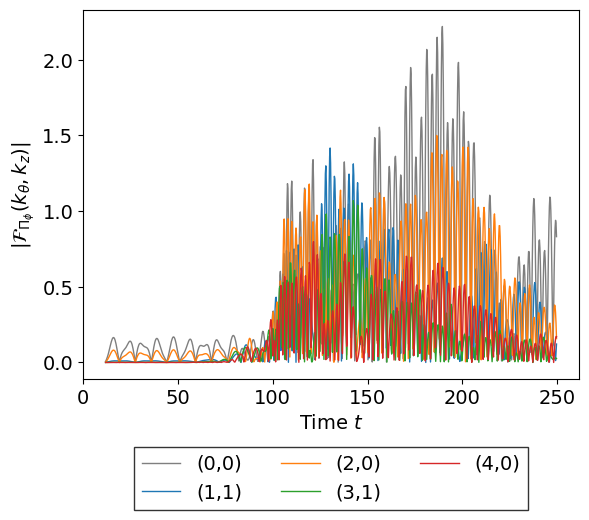}
    \caption{Absolute value of the $\{mn\} = \{0\,0\}, \{1\,1\}, \{2\,0\}, \{3\,1\},$ and $\{4\,0\}$ Fourier modes of the massive radiation $\Pi_\phi$ from a $\lambda=1$ string with initial amplitude $A_0=4$, measured on a cylinder at $R=64$.  The propagating radiation modes are $\{pmn\} = \{6\, 0\, 0\}$, $\{6\, 1\, 1\}$, $\{6\, 2\, 0\}$, $\{6\, 3\, 1\}$, and $\{6\, 4\, 0\}$. We note also the initial presence of oscillating self-fields, $\{1\, 1\, 1\}$, $\{2\, 0\, 0\}$ and $\{2\, 2\, 0\}$.}
    \label{amp1massivemodes}
\end{figure}

\subsection{Massive Radiation Analysis}\label{massiveanalysis}

In this section, we present a quantitative analysis of the massive radiation from oscillating string configurations. Simulations are set up as outlined in Section \ref{simulationsetup}, and we bear in mind the convergence tests and discussion in Section \ref{convergencetesting}.

\begin{figure}
    \centering
    \includegraphics[width=\linewidth]{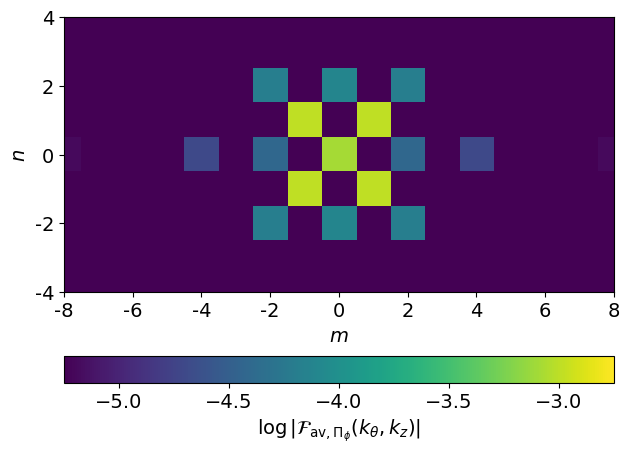}    
    \includegraphics[width=\linewidth]{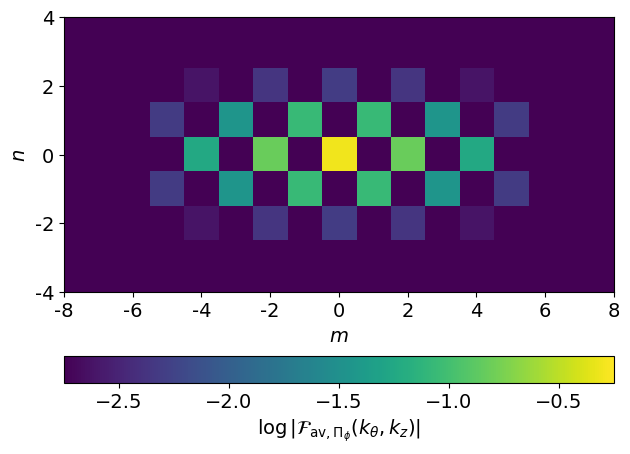}
    \includegraphics[width=\linewidth]{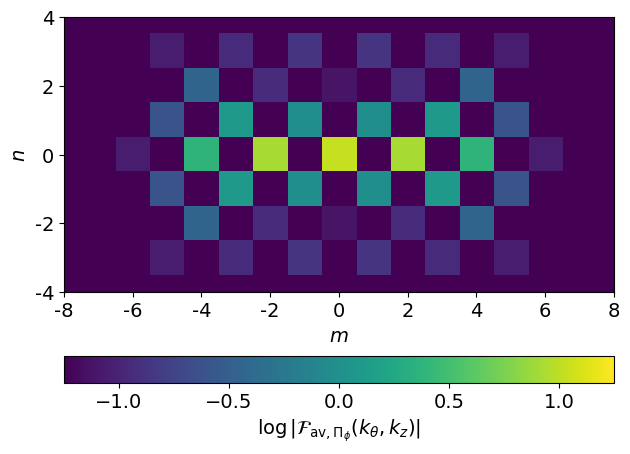}
    \caption{2D Fourier eigenmodes of the massive radiation $\Pi_\phi$ from a $\lambda=1$ string at late time $t=160$, measured on a cylinder at $R=64$ with $\sigma=0$ and time averaged over approximate half-period $\Delta t/2 = 66/4$. The horizontal axis is the angular eigenvalue $m$, while the vertical is the $z$-dependent wavenumber $n$. The top figure is for an initial amplitude $A_0=1$, the middle is for intermediate $A_0=4$ and the bottom is large $A_0=8$, showing an increasing trend of higher harmonics and a significant increase in amplitude, highlighted by the changing scales.}
    \label{checkerboardmassive4}
\end{figure}

\begin{figure}[t]
    \centering
    \includegraphics[width=\linewidth, trim=0 0 0 0, clip]{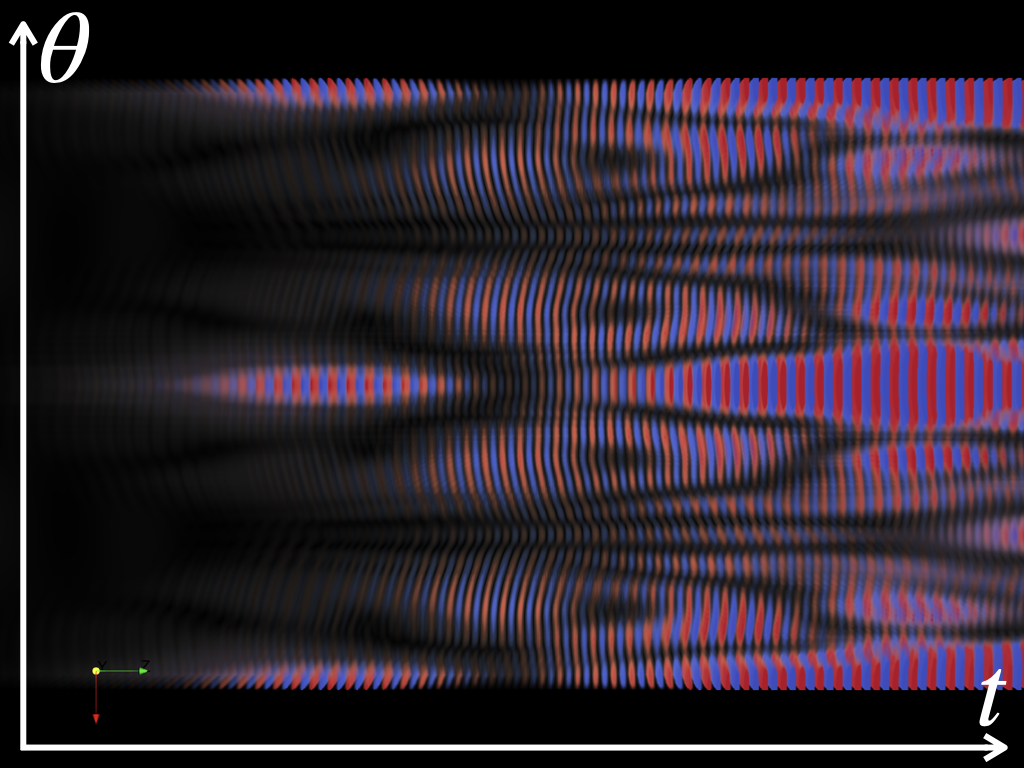}
    \caption{Volume rendering in spacetime $(t, \theta, z)$ of the massive radiation $\Pi_\phi$ from a $\lambda = 10$ string with initial amplitude $A_0 = 1$ over time, measured on a cylinder at $R = 64$. The time axis runs left to right, the azimuthal angle $\theta$ from bottom to top and the $z$-axis out of the page. Complex resonant patterns characterise the radiation.}
    \label{ParaviewMassiveScreenshotAmp1}
\end{figure}

We begin by presenting a detailed investigation of the massive radiation from $\lambda=1$ and $\lambda=10$ strings with small amplitude $A_0=1$ ($\varepsilon=0.20$) and larger amplitudes $A_0=4$ and $A_0=8$ ($\varepsilon=0.68\;\rm{and}\;1$). We perform quantitative analysis by extracting and Fourier decomposing the massive radiation field $\Pi_\phi$, defined by equation (\ref{massivediagnostic}), on a diagnostic cylinder at fixed radius $R=64$.

We also perform a more detailed scan over $0.3\leq\lambda\leq2$, for $\lambda$ spaced by $\Delta\lambda = 0.1$. We determine the $\lambda$-dependence of the massive spectrum, including the primary radiation modes and energy loss. For this finely spaced scan, we concentrate primarily on two relative amplitudes; $A_{\,\rm{rel}} = 0.5$ using $A_0=4$ with $L=32$ ($\varepsilon=0.68$) and $A_{\,\rm{rel}} = 0.875$ using $A_0=3.5$ with $L=16$ ($\varepsilon=0.96$), where $A_{\,\rm{rel}}$ is defined by (\ref{rescale}).

To clearly demonstrate the qualitatively different nature of massive radiation from massless radiation, we first visualise the massive diagnostic $\Pi_\phi$ in three dimensions. Taking $\lambda=1$ and the intermediate amplitude $A_0=4$ as a representative example, the signal is illustrated in Figure~\ref{ParaviewMassiveScreenshotAmp4}. Although the radiation is predominantly dipole, the spectrum is significantly more complex than the massless quadrupole radiation from the same configuration (see \cite{Drew2019}). This is particularly evident in animations. We also observe that the different phase and group velocities lead to short wavelength modes travelling rapidly forward within larger, slower-moving outgoing wavepackets.

\begin{figure}[!ht]
    \centering
    \includegraphics[trim = 0 2cm 0 0, clip, width=\linewidth]{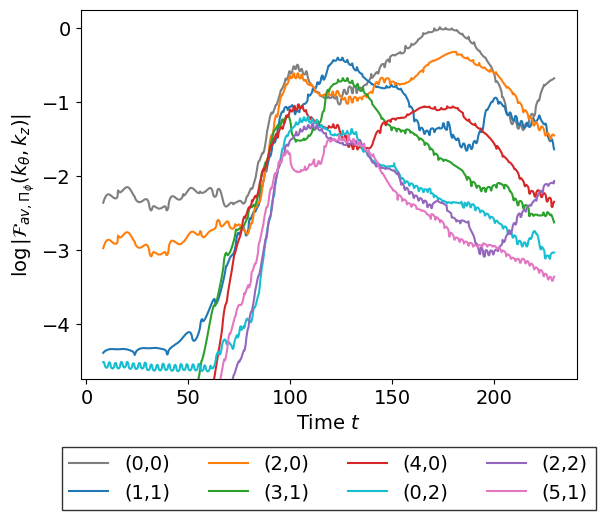}
    \includegraphics[width=\linewidth]{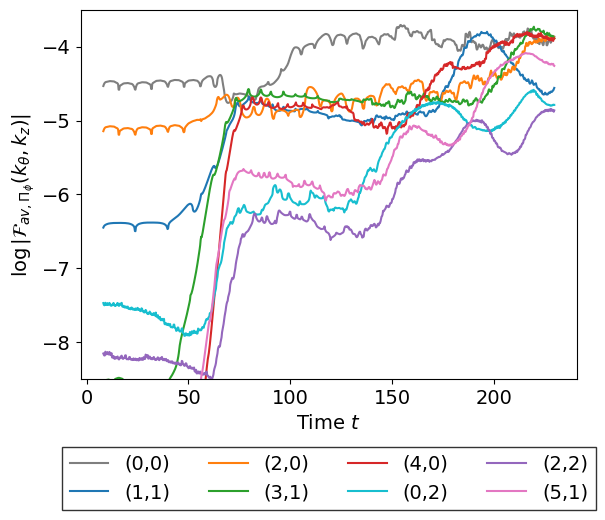}
    \caption{Dominant 2D Fourier modes of the massive radiation $\Pi_\phi$ from a $\lambda=1$ (top) and $\lambda=10$ (bottom) string with initial amplitude $A_0=4$, measured on a cylinder at $R=64$ and time averaged over approximate half-period $\Delta t/2 = 66/4$.}
    \label{massivemagmodes1} 
\end{figure}    

\begin{figure}[!ht]
    \centering
    \includegraphics[trim = 0 2cm 0 0, clip, width=\linewidth]{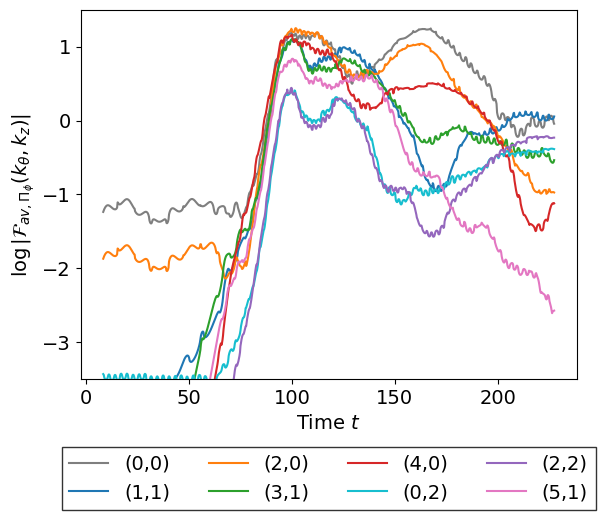}
    \includegraphics[width=\linewidth]{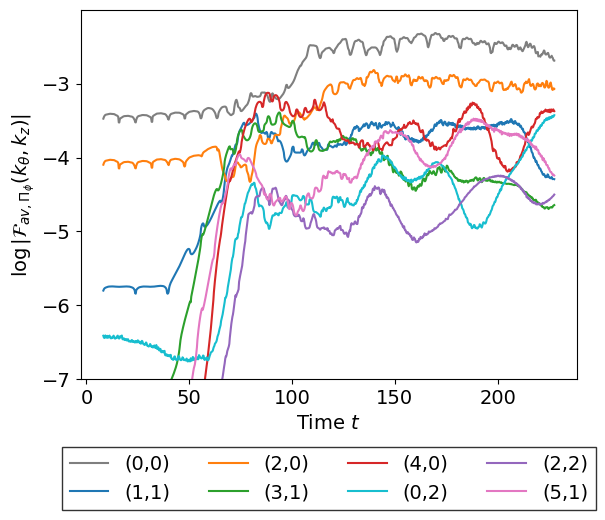}
    \caption{Dominant 2D Fourier modes of the massive radiation $\Pi_\phi$ from a $\lambda=1$ (top) and $\lambda=10$ (bottom) string with initial amplitude $A_0=8$, measured on a cylinder at $R=64$ and time averaged over approximate half-period $\Delta t/2 = 66/4$.}
    \label{massivemagmodes2} 
\end{figure}   

\subsubsection{Mode Decomposition}

Here, we undertake a Fourier analysis of the massive radiation signal $\Pi_\phi$ on the diagnostic cylinder at $R=64$ to quantify the effects described above. The time evolution of the largest amplitude eigenmodes is plotted in Figure~\ref{amp1massivemodes} for $\lambda=1$ and $A_0=4$, where the individual modes are obtained using a 2D Fast Fourier Transform (FFT). We recall from Section \ref{analyticexpectations} that the $p$ (time) eigenvalue determines whether or not a certain mode will propagate. Measuring the time-dependence of the signal, we identify the massive propagating modes $\{6\, 0\, 0\}$, $\{6\, 1\, 1\}$, $\{6\, 2\, 0\}$,  $\{6\, 3\, 1\}$ and $\{6\, 4\, 0\}$. This $p=6$ time-dependence is consistent with the requirement that the frequency be above the (\ref{massthresholdlow}) mass threshold, given approximately by $p > p_{\rm min} \approx   \sqrt \lambda \,/\, (2\pi/L) \approx 5.1 $ for $L=32$. We also identify the long-range self-field excitations $\{1\, 1\, 1\}$, $\{2\, 0\, 0\}$ and $\{2\, 2\, 0\}$ sourced by the massless self-field (i.e. before the propagating modes reach the analysis cylinder), as discussed in Section \ref{selffieldsection}. This means that, despite having the appearance of a simple dipole in Figure~\ref{ParaviewMassiveScreenshotAmp4}, the radiation signal is in fact more complex, with monopole and quadrupole modes also present at comparable magnitude. One explanation for this apparent difference is that the radiation pattern is somewhat `beamed,' requiring a combination of modes to achieve angular localisation in comparison with the pure $\{1\,1\,1\}$ dipole. Finally, we observe that the radiation propagation velocity $v_{\rm g}\approx 0.5$, measured from the first arrival of the propagating signal at the cylinder, agrees with the predicted $v_{\rm g}=0.48$.

%\ad{This graph is sigma=1. Sigma=0 graph looks similar, overall higher amplitude around 1.5x, slightly different distribution of modes but not significant.}

\begin{figure}
    \centering
    \includegraphics[width=\linewidth]{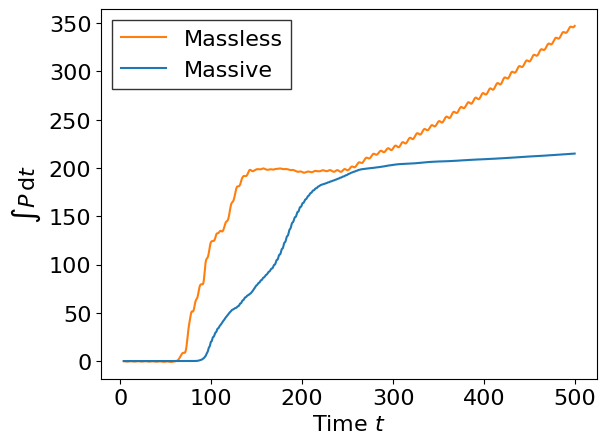}
    \includegraphics[width=\linewidth]{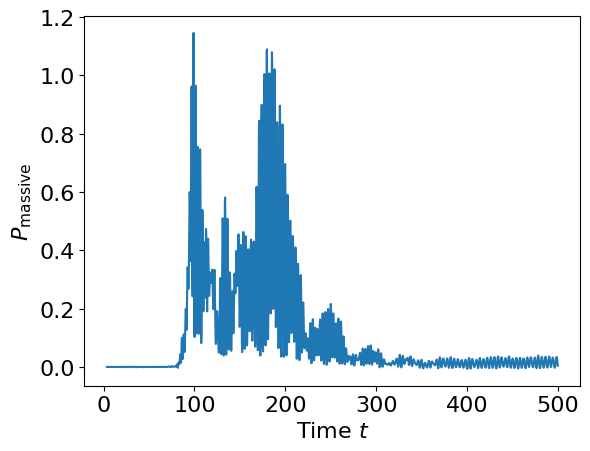}
    \includegraphics[width=\linewidth]{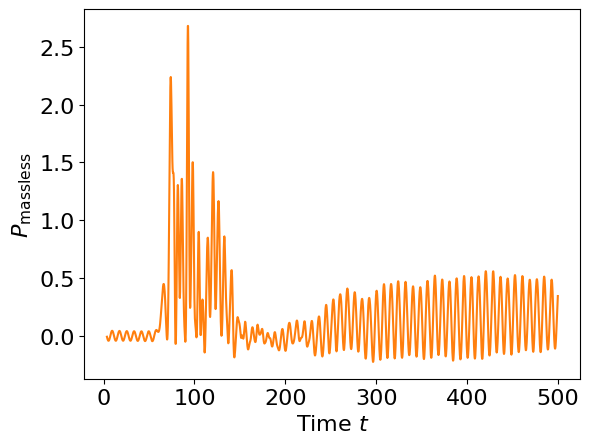}   
    \caption{Massive and massless radiation emitted from a $\lambda=0.8$ string with amplitude $A_0=6$ and $L=16$, giving the effective relative amplitude $A_\mathrm{rel} \,(\sim \varepsilon_\mathrm{eff}) = 1.5$, a highly relativistic configuration. The top graph shows the cumulative integrated massive and massless components of the `Poynting' vector, $P_{\rm{massive}}$ \eqref{Pmassive} and $P_{\rm{massless}}$ \eqref{Pmassless}, on the diagnostic cylinder at $R=64$ over time. The middle and bottom graphs show the massive and massless components respectively integrated over the diagnostic cylinder and plotted over time.}
    \label{energycompeting}
\end{figure}

Having investigated the radiation from $A_0=4$, we further analyse $A_0 = 1$ and $A_0 = 8$ for $\lambda=1$ to determine the dependence on amplitude of high-order harmonics. Figure~\ref{checkerboardmassive4} shows the time-averaged magnitude of all measured eigenmodes for $A_0 = 1, \,4$ and $8$ extracted on the cylinder at $t=160$. The time-average is calculated by extracting each separate Fourier mode $\mathcal{F}_{\Pi_\phi}(k_\theta,k_z)$ and averaging over time, using
\begin{equation}\label{modes}
\mathcal{F}_{\mathrm{av}, \Pi_\phi}(k_\theta,k_z) = \sum^{t = \Delta t/4}_{t = -\Delta t/4}{2\,\mathcal{F}_{\Pi_\phi}(k_\theta,k_z)} / \Delta t\,,
\end{equation}
where $\Delta t$ is approximately one period of oscillation. We observe that, at small amplitude, primarily dipole and quadrupole radiation and self-field modes are measured. As the configurations probe higher (nonlinear) amplitudes as $\varepsilon \rightarrow 1$, higher frequency modes become activated and a checkerboard pattern emerges. An $m+n$ {\it even} selection rule applies as for the massless radiation modes \cite{Drew2019}, although the distribution of massive modes is more constrained in the $z$-direction. There is also a more nonlinear dependence of the total magnitude on the initial amplitude $A_0$, as can be observed from the different logarithmic scales required to plot each case.

The additional complexity and challenge of higher order massive radiation is further illustrated by considering the spectrum for larger $\lambda$. The radiation pattern shown in Figure~\ref{ParaviewMassiveScreenshotAmp1} is emitted by low amplitude $A_0=1$ for $\lambda=10$. Overall, the signal has a significantly smaller magnitude and a higher $p_{\rm{min}}$ than both the massless radiation and the massive radiation for $\lambda=1$. This is because, unlike massless radiation which is independent of $\lambda$ to leading order, massive radiation becomes more strongly suppressed as $\lambda$ increases. This will be discussed in detail in Section \ref{expdep}. The signal begins as an isolated dipole with a $\{17\, 1\, 1\}$ mode contribution ($p_{\rm min}\approx16.1$). At late time, we begin to observe resonant effects introducing higher angular harmonics including $m=2,\, 3,\, \mathrm{and} \, 4$, which interchange amplitudes and generally increase during the simulation, with the quadrupole mode $\{17\, 2\, 0\}$ becoming comparable in magnitude to the dipole. Not only does the signal evolve between harmonics, the varying amplitude offers indications of stimulated emission through string-radiation interactions. An important caveat here at this low amplitude and high $\lambda$, however, is that this massive radiation signal becomes more susceptible to numerical effects, especially those discussed previously for AMR in section \ref{convergencetesting}.  For this reason, Figure~\ref{ParaviewMassiveScreenshotAmp1} should be interpreted as a qualitative insight into the complexity of massive radiation, rather than an accurate physical solution.

%Due to the overall small magnitude of the radiation and the interpolation between refinement levels, this behaviour may not necessarily be robust to changes in the adaptive mesh, meaning that aspects of the spectrum may be unphysical. 

%However, in this case the massive amplitude is $\sim 100\times$ smaller than the corresponding massless radiation signal, so has a negligible effect on string motion.

\begin{table}
\centering
\caption{$\lambda$-dependence of $p_{\mathrm{min}}$ for the dipole ($n=1$) and quadrupole ($n=0$) Fourier modes for strings with initial amplitude $A_0=4$ and wavelength $L=32$, characterised by $A_{\,\rm{rel}} = 0.5$, and initial amplitude $A_0=3.5$ and wavelength $L=16$ with $A_{\,\rm{rel}} = 0.875$. Two models for $\alpha$ are considered, the Nambu-Goto model $\alpha_{\rm{NG}}$ calculated using equation (\ref{alpha}) and the sinusoidal model $\alpha_{\sin}$ using equation (\ref{sinusoidal}). Strings radiate primarily into the Fourier mode $p_{\rm{min}}$ when $\lambda < \lambda_{pn}$, its corresponding threshold. As $\lambda$ is increased, the value of $p_{\mathrm{min}}$ also increases, so lower frequency modes become unavailable.}
\centering
\vspace{0.5cm}
\begin{ruledtabular}
\begin{tabular}{l c c c c}
$A_{\mathrm{rel}} = 0.5$ \vspace{0.1cm} & & & & \\
\vspace{0.1cm} $p_{\mathrm{min}}$ & \multicolumn{4}{c}{$\lambda_{pn}$}\\
& \multicolumn{2}{c}{$\alpha_{\rm NG}=1.15$} & \multicolumn{2}{c}{$\alpha_{\rm sin}=1.14$} \\
&  $n=1$ & $n=0$ & $n=1$ & $n=0$ \\

3 & 0.224 & 0.262 & 0.228 & 0.267 \\

4 & 0.428 & 0.466 & 0.436 & 0.475 \\

5 & 0.690 & 0.729 & 0.703 & 0.742 \\

6 & 1.011 & 1.049 & 1.029 & 1.068 \\

7 & 1.390 & 1.428 & 1.415 & 1.454 \\

8 & 1.827 & 1.866 & 1.860 & 1.899 \\

9 & 2.323 & 2.361 & 2.364 & 2.403 \\

\vspace{-0.35cm} & & & & \\

\hline\hline \vspace{-0.35cm} & & & & \\
$A_{\mathrm{rel}} = 0.875$ \vspace{0.1cm} & & & & \\
\vspace{0.1cm} $p_{\mathrm{min}}$ & \multicolumn{4}{c}{$\lambda_{pn}$}\\

& \multicolumn{2}{c}{$\alpha_{\rm NG}=1.437$} & \multicolumn{2}{c}{$\alpha_{\rm sin}=1.373$} \\ 
& $n=1$ & $n=0$ & $n=1$ & $n=0$ \\
\midrule
2 & 0.145 & 0.299 & 0.173 & 0.327 \\

3 & 0.518 & 0.672 & 0.582 & 0.736 \\

4 & 1.041 & 1.195 & 1.155 & 1.309 \\

5 & 1.713 & 1.867 & 1.891 & 2.045 \\

6 & 2.534 & 2.688 & 2.791 & 2.945 \\

7 & 3.505 & 3.659 & 3.854 & 4.008 \\

\end{tabular}
\end{ruledtabular}
\label{L32thresholds}

\end{table}

\begin{figure}
    \centering
    \includegraphics[trim={0 2cm 0 0},clip, width=\linewidth]{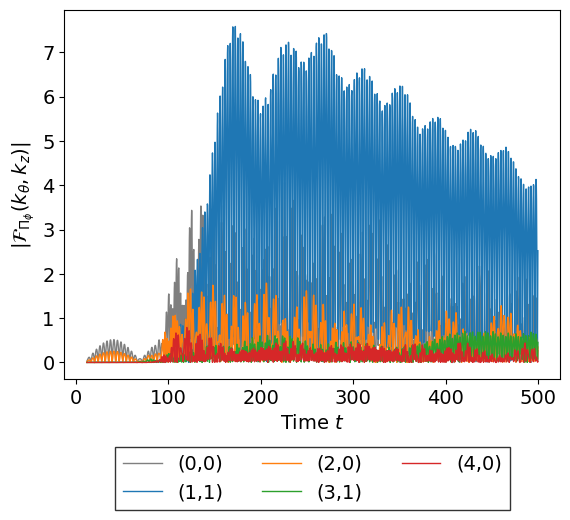}
    \includegraphics[width=\linewidth]{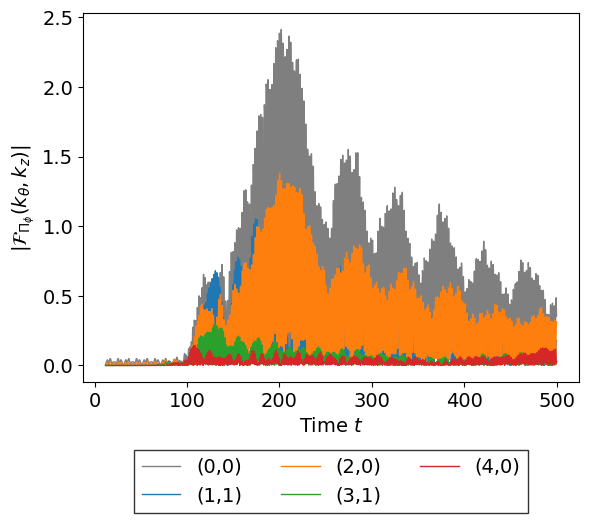}    
   \caption{Absolute value of the $\{mn\} = \{0\,0\}, \{1\,1\}, \{2\,0\}, \{3\,1\},$ and $\{4\,0\}$ Fourier modes of the massive radiation $\Pi_\phi$ from strings with $\varepsilon=0.875$, measured on a cylinder at $R=64$ for $\lambda=0.8$ (top) and $\lambda=2$ (bottom). Although dipole radiation usually dominates, as in the $\lambda=0.8$ case, the mass threshold $\lambda_{p0}$ for the zero-mode and quadrupole radiation for a given $p$ is higher than for dipole modes (i.e. easier to satisfy $\lambda < \lambda_{pn}$), as shown in Table \ref{L32thresholds}. The zero and quadrupole modes can therefore be dominant in some tuned cases, as shown for $\lambda=2$.}
    \label{A3.5L16}
\end{figure}

\subsubsection{Relative Energy Loss to Massive and Massless Modes}\label{relenergy}

In this subsection, we make a quantitative comparison between the magnitude of the dominant massive and massless modes for different values of $\lambda$ and $A_0$. We first compare the magnitude of energy emitted via massive radiation for $\lambda=1$ and $\lambda=10$. Figures \ref{massivemagmodes1} and \ref{massivemagmodes2} show the time-average for the eight strongest massive modes for $\lambda=1$ and $\lambda=10$ strings, with initial amplitudes $A_0=4$ and $A_0=8$ respectively. (The $\lambda=1$ results plot the time dependence of the modes plotted in the lower two panels of Figure \ref{checkerboardmassive4}.) We observe for each amplitude that there is a difference in scale of $\sim 10^4\times$ between the magnitude of the most dominant modes for $\lambda=1$ and $\lambda=10$. The $\lambda=10$ radiation is therefore heavily suppressed, as predicted by the significant increase in mass threshold.

Comparing to the massless radiation in Figure 20 of \cite{Drew2019}, the magnitude of the massive modes for $\lambda=1$ and $A_0=4$ is also $\sim 1000\times$ smaller than the massless modes, meaning that radiation via massive radiation can effectively be taken to be negligible. This agrees with observations made in \cite{Drew2019}, where for $\lambda=1$ and $A_0=4$, the massive radiation was so negligible as not to be noticeable as a contribution to the total energy loss. This ratio is even more extreme for $\lambda=10$, where we observe massive radiation to be $\sim 10^7\times$ smaller in magnitude than massless radiation.\footnote{We recall from the discussion in \cite{Drew2019} that the massless radiation spectrum for $\lambda=10$ is very similar to the spectrum for $\lambda=1$.} This comparison for $A_0=8$ is less extreme, where we have a factor of only $\sim 100\times$ between the massless and massive modes for $\lambda=1$. This is due to the fact that the higher initial amplitude corresponds to more relativistic string oscillations and larger accelerations, allowing massive modes to be activated more easily.

Although the massive radiation is typically negligible as an energy loss mechanism for the configurations described above, it is possible for energy loss from massive modes to compete with massless modes for very relativistic configurations with curvature comparable to the string width, i.e. in the limit with low $\lambda$ and high $\varepsilon$. This makes it easier to activate massive modes, due to the lower mass threshold and more relativistic motion. Figure \ref{energycompeting} shows the massive and massless components of the `Poynting' diagnostic, $P_{\rm{massive}}$ \eqref{Pmassive} and $P_{\rm{massless}}$ \eqref{Pmassless}, integrated over the diagnostic cylinder for a highly relativistic $\lambda=0.8$ string with $A_0=6$ and $z$-dimension $L=16$. We observe that the energy emitted via massive radiation is of a comparable magnitude to the massless channel, particularly after the initial burst.  This string configuration has such a large amplitude ($\varepsilon_{\mathrm{eff}} > 1$), that extended regions of the string contract and coalesce as they approach the speed of light for a protracted time (so-called relativistic string `lumps', unlike momentary point-like `cusps'). These highly relativistic and degenerate string regions disintegrate (essentially self-annihilate) into beamed radiation into all the available massless and massive channels.  However, this only occurs on the first large amplitude ($\varepsilon_{\mathrm{eff}} > 1$) oscillation, after which massless radiation once again strongly dominates for subsequent ($\varepsilon < 1$) oscillations.  This nonlinear phenomenon is likely to be relevant for high curvature regions in network simulations, particularly those using fixed comoving width which have a small effective $\lambda$.

\begin{figure*}
    \centering
    \includegraphics[width=0.48\linewidth]{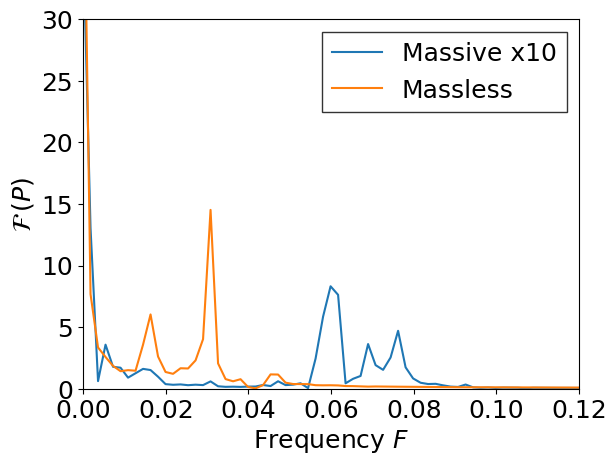}
    \includegraphics[width=0.48\linewidth]{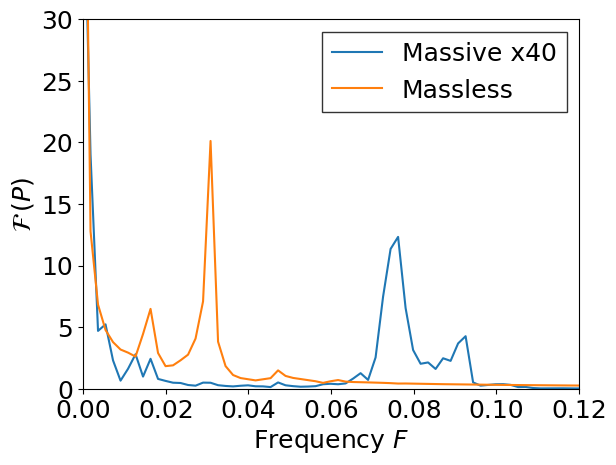}    
    \includegraphics[width=0.48\linewidth]{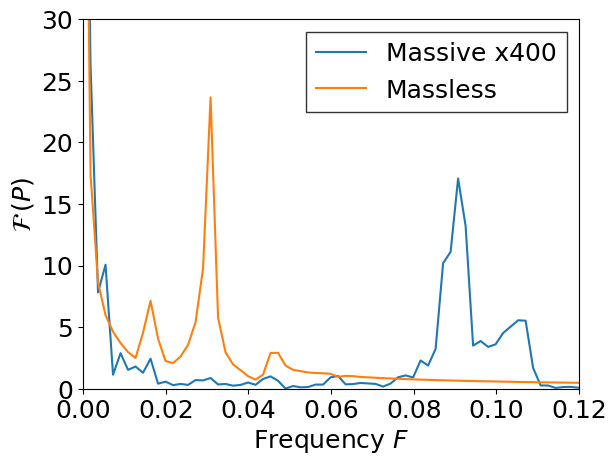}    
    \includegraphics[width=0.48\linewidth]{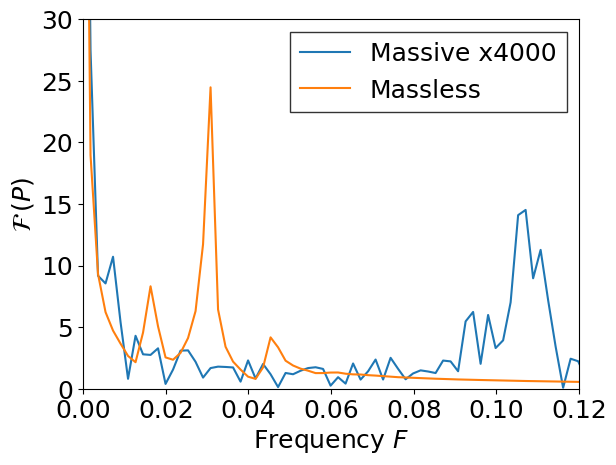}    
    \caption{Fourier mode decomposition of $P_{\mathrm{massive}}$ \eqref{Pmassive} and $P_{\mathrm{massless}}$ \eqref{Pmassless} for $\varepsilon=0.5$. We plot $\lambda=0.4$ (top left), $\lambda=0.6$ (top right), $\lambda=0.9$ (bottom left) and $\lambda=1.2$ (bottom right). The massive signal has been integrated from $t=90$ to $228$ to capture the initial burst of radiation, whilst minimising effects of radiation reflected from the boundaries. Note the change in scale of the massive radiation for each $\lambda$ by a ratio indicated in the legend.}
    \label{FFTextraction}
\end{figure*}

\subsubsection{Radiation Harmonics and $\lambda$-Dependence}

In this subsection, we scan over a range of $0.3\leq\lambda\leq2$ to determine the more detailed $\lambda$-dependence of the massive spectrum. We concentrate primarily on two amplitudes; $A_{\,\rm{rel}}=0.5$ ($A_0=4$ with $L=32$) and $A_{\,\rm{rel}} = 0.875$ ($A_0=3.5$ with $L=16$).
 
Using the relationship derived in equation (\ref{lambdathresholds}), we calculate the $\lambda_{pn}$ threshold values at which the lowest $p$ harmonic, $p_{\mathrm{min}}$, that can be activated for a certain amplitude changes. These values are presented in Table \ref{L32thresholds} for $A_{\,\rm{rel}}=0.5$ and $A_{\,\rm{rel}}=0.875$. We consider two models for the fractional increased path length $\alpha$; the Nambu-Goto model $\alpha_{\rm{NG}}$ calculated numerically using equation (\ref{alpha}) and the sinusoidal model $\alpha_{\sin}$ using equation (\ref{sinusoidal}). For each model, we obtain different $\lambda_{pn}$ values for each $p_{\mathrm{min}}$ and for different values of $n$. Taking a $\lambda=1.8$ string with $A_{\,\rm{rel}}=0.875$ as an example, from the Nambu-Goto model we obtain $\alpha_{\rm NG} = 1.437$, which gives for the dipole $(n=1)$ mode $p_{\mathrm{min}}=6$, but $p_{\mathrm{min}}=5$ for the quadrupole $(n=0)$ mode. For the same configuration using the sinusoidal model, we obtain $\alpha_{\rm sin}=1.373$, which gives $p_{\mathrm{min}}=5$ for both the $n=1$ and $n=0$ modes. This means that in practice, there are a range of potential $\lambda_{pn}$ for each $p_{\mathrm{min}}$ due both to the theoretical uncertainty about the appropriate $\alpha$ model and the range of available $n$ modes. An example of a change of the dominant radiative mode is presented in Figure \ref{A3.5L16}, which shows the Fourier mode decomposition for $\lambda=0.8 \; \mathrm{and} \; 2.0$, again for $A_{\,\rm{rel}} = 0.875$. We clearly observe that the Fourier decomposition of the radiation changes depending on $\lambda$; $\lambda=0.8$ radiates primarily in the $\{1\,1\}$ dipole mode, but $\lambda=2$ primarily in the $\{0\,0\}$ zero mode and $\{2\,0\}$ quadrupole mode. This corresponds with the thresholds in Table \ref{L32thresholds}; $\lambda=0.8$ lies below the dipole threshold values for $p_{\mathrm{min}}$ for all models, whereas $\lambda=2$ lies between the dipole and quadrupole thresholds for the sinusoidal model for $p_{\mathrm{min}}=5$. We note that the use of the string wavelength $L=16$ for $A_{\,\rm{rel}} = 0.875$ in Table \ref{L32thresholds} provides easier access to lower $p_{\mathrm{min}}$ compared with $L=32$, as well as a larger difference in $\lambda_{pn}$ between the dipole and quadrupole thresholds. We also observe, as expected, that the frequency of the modes increases and the magnitude of the radiation decreases as $\lambda$ increases.

%The left column of Figure \ref{FFTextraction} shows the massive part of the Poynting diagnostic from equation (\ref{momentumdiag}), ${\Pi}_\phi {\cal D} \phi$, integrated over the diagnostic cylinder for the chosen $\lambda$ values and plotted over time from $0 \le t \leq 500$. We first observe that the frequency of the radiation increases as $\lambda$ increases, although this is difficult to see directly from the data due to the high frequencies. We further note that the maximum magnitude of the radiation decreases as $\lambda$ increases, as expected from the increasing mass threshold.

In order to test the accuracy of the $\lambda_{pn}$ threshold predictions, we qualitatively analyse the emitted massive spectra from the $A_{\,\rm{rel}}=0.5$ and $A_{\,\rm{rel}}=0.875$ configurations. Figure \ref{FFTextraction} shows results obtained for $A_{\,\rm{rel}}=0.5$, where the $\lambda$ values plotted have been chosen to lie between the $p_{\rm{min}}$ threshold values $\lambda_{pn}$ in Table \ref{L32thresholds}; $\lambda=0.4,\, 0.6,\, 0.9 \; \mathrm{and} \; 1.2$. We perform Fourier transforms on the extracted signals, $P_{\mathrm{massive}}$ \eqref{Pmassive} and, for comparison, $P_{\mathrm{massless}}$ \eqref{Pmassless}, choosing to integrate from $90 \leq t \leq 228$. This is not a straightforward choice as, observing the massive diagnostic over time, we see that the length of the initial signal varies significantly and in some cases unpredictably between $\lambda$. We judge by the extracted signals that integrating from $90 \leq t \leq 228$ is sufficient to capture the initial burst of radiation. We also note that this is equivalent to approximately four periods of oscillation of the strings.

\begin{figure}[t!]
    \centering
    \includegraphics[width=\linewidth]{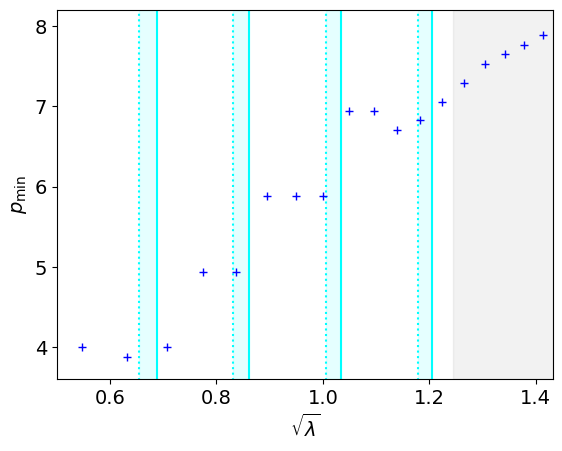}    
    \includegraphics[width=\linewidth]{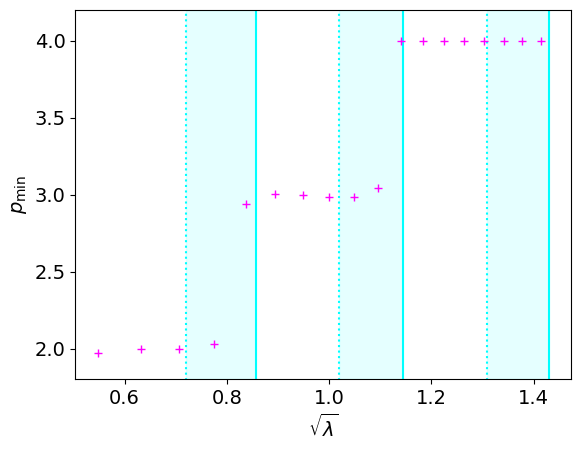}    
    \caption{Measured $p_{\mathrm{min}}$ as a function of $\sqrt{\lambda}$ for massive radiation from string configurations with $A_{\,\rm{rel}}=0.5$ (top) and $A_{\,\rm{rel}}=0.875$ (bottom). The graphs each summarise data from approximately twenty simulations. Shaded regions in cyan show the predicted thresholds $\lambda_{pn}$ from Table \ref{L32thresholds}, where the shading encompasses the dependence on $\alpha$ and $n$; the solid cyan line represents the highest predicted $\lambda_{pn}$, coming from the $\alpha_{\rm sin }$ model with $n=0$ for each $p$ and the dashed cyan line is the lowest reasonable $\lambda_{pn}$, coming from $\alpha_{\rm NG}$ with $n=1$. In the top plot, we observe four clear harmonic suppression thresholds before the radiation becomes so weak that it becomes dominated by numerical effects (grey shaded region). In the bottom plot, we observe three clear thresholds.}
    \label{scattergraphs}
\end{figure}

The Fourier transforms in Figure \ref{FFTextraction} provide a very clear picture of the mode decomposition of the massive signal for $A_{\,\rm{rel}}=0.5$ and $L=32$. We first note that, as expected, the overall magnitude of the massive radiation component decreases by orders of magnitude as $\lambda$ increases, whilst the massless signal increases slightly due to reduced radiation backreaction.\footnote{Some additional Kreiss-Oliger damping has been applied to the higher frequency modes to make the lowest propagating mode clearer. This does not change the $p$ value of the modes.} The massless signal in each case radiates primarily into the $p=2$ harmonic, as determined in \cite{Drew2019}, along with a smaller $p=1$ signal. This provides a very clear benchmark against which the massive signals can be compared. We clearly observe the increase in the massive $p_{\mathrm{min}}$ as $\lambda$ increases, and can deduce the values by comparison with the massless peak. In order to be consistent with the model outlined in Section \ref{massive}, we expect $p_{\mathrm{min}}$ to take integer values, increasing stepwise as the minimum radiative mode increases with $\lambda$. We observe $p_{\mathrm{min}} = 4, 5, 6$ and $7$ for $\lambda=0.4,\, 0.6,\, 0.9 \; \mathrm{and} \; 1.2$ respectively, agreeing with the predicted $\lambda_{pn}$ presented in Table \ref{L32thresholds}. We further note that the massive peak for the lowest massive harmonic is not always a clean signal, sometimes comprising of a double peak. This is consistent with the different $\lambda_{pn}$ predicted for different $n$ harmonics.

The top panel of Figure \ref{scattergraphs} shows the measured dominant massive harmonic $p_{\mathrm{min}}$ normalised against the quadrupole $p=2$ massless harmonic for $0.3\leq\lambda\leq2$ spaced by $\Delta\lambda\approx 0.1$, for $A_{\,\rm{rel}}=0.5$. Measured $p_{\mathrm{min}}$ values are obtained by numerically extracting the position of the peak of the Fourier transform of the massive signal from Figure \ref{FFTextraction}. We indicate the predicted thresholds $\lambda_{pn}$ from Table \ref{L32thresholds} using cyan shaded regions to encompass the dependence on the $\alpha$ model and $n$ harmonic. We observe the presence of distinct harmonic thresholds as predicted in Section \ref{massive} corresponding to the predicted integer values of $p_{\mathrm{min}}$ as expected until $\lambda \gtrsim 1.5$, where the levels become less distinct and merge together. This provides strong evidence for the underlying mechanism for radiation into massive modes being via higher harmonic excitations of the fundamental mode of string oscillation for low $\lambda$. We also note that the thresholds correspond more closely with the sinusoidal model of the path length than the Nambu-Goto model.

%\begin{figure}[t!]
%    \centering
%    \includegraphics[width=\linewidth]{L=16massive_levels_thresholds_sqrt_correct_damping.png}
%    \caption{Measured $p_{\mathrm{min}}$ as a function of $\sqrt{\lambda}$ for massive radiation from string configurations with $A_{\,\rm{rel}}=0.875$. The graph summarises data from approximately twenty simulations. Shaded regions in cyan show the predicted thresholds $\lambda_{pn}$ from Table \ref{L32thresholds}, where the shading encompasses the dependence on $\alpha$ and $n$; the solid cyan line represents the highest predicted $\lambda_{pn}$, coming from the $\alpha_{\rm sin }$ model with $n=0$ for each $p$ and the dashed cyan line is the lowest reasonable $\lambda_{pn}$, coming from $\alpha_{\rm NG}$ with $n=1$. Here, we observe four clear harmonic radiation thresholds which do not correspond with integer harmonics.}
 %   \label{scattergraphs2}
%\end{figure}

The bottom panel of Figure \ref{scattergraphs} shows $p_{\mathrm{min}}$ plotted against $\sqrt{\lambda}$, determined using the same method as above, for the highly non-linear regime with $A_{\,\rm{rel}} = 0.875$. We observe qualitatively the same behaviour as for $A_{\,\rm{rel}}=0.5$, namely that distinct thresholds are present, in this case for the full range of $\lambda$ values plotted. Again, these thresholds correspond well with the integer values of $p_{\mathrm{min}}$ predicted in Table \ref{L32thresholds}. This demonstrates that, although the $\lambda_{pn}$ model is derived for low relative amplitude, the mode predictions still apply as $\varepsilon \rightarrow 1$. Furthermore, the higher magnitude of radiation emitted by this more relativistic configuration results in clear harmonics being radiated up to higher $\lambda$ than for $A_{\,\rm{rel}} = 0.5$, as numerical effects are not yet large enough to interfere with the physical radiation.

%\begin{figure}
%    \centering
%    \includegraphics[width=0.9\textwidth]{Chapter4/L=32massive_loglog2.png}
%    \includegraphics[width=0.9\textwidth]{Chapter4/L=32massive_loglinear2.png}
%    \caption{Log-log (top) and log-linear (bottom) plots of the massive radiation diagnostic $\Pi_\phi$ integrated over a diagnostic cylinder at $R=64$ from $90 \leq t\leq 228$ (labelled $P_{massive}$) for $\alpha = 1.15$ and a range of $0.3 \leq \lambda \leq 2.5$. Conclusions? (lambda=1 is a bug, needs redoing)}
%    \label{L32_massive_summary}
%\end{figure}

\subsubsection{$\lambda$-dependence of Massive Radiation Spectrum}\label{expdep}

%Massive radiation from a sinusoidally displaced string is predicted to be suppressed exponentially as the wavelength increases. A derivation of this result is given in \cite{Olum2000}, which predicts that the total energy $E$ emitted by an oscillating sinusoidal string for a fixed relative amplitude $\varepsilon$ and fixed mass $\lambda$ is given by
%\begin{equation}
%    E(L) = \sqrt{L}e^{-\beta L/(2\pi^2)}\,,
%\end{equation}
%where $\beta$ is a positive constant and $L$ is the wavelength, equivalent to the $z$-dimension of the simulation box. This model can be modified to obtain an expression in terms of the mass parameter $\lambda$ as follows:
%\begin{equation}\label{newenergy}
%    E(L,\varepsilon,\lambda) = \sqrt{\frac{L}{\varepsilon}}\exp{\left[-\gamma\, \sqrt{\lambda}\,\frac{L}{\varepsilon}\frac{1}{2\pi^2}\right]}\,,
%\end{equation}
%where $\gamma$ is a positive constant and we have added back in the dependence on relative amplitude $\varepsilon$. 

%The identification of distinct harmonics in the previous subsection indicates that the mechanisms generating massive radiation are perturbative. 

In this subsection, we investigate the overall $\lambda$-dependence of the massive radiation spectrum. We again concentrate primarily on two relative amplitudes; $A_{\,\rm{rel}}=0.5$ ($A_0=4$ with $L=32$) and $A_{\,\rm{rel}} = 0.875$ ($A_0=3.5$ with $L=16$). 

We have shown in the previous subsection that, for $A_{\,\rm{rel}}=0.5$ with $\lambda \lesssim 1.5$ and for all $\lambda$ studied for $A_{\,\rm{rel}}=0.875$, massive radiation is only emitted in harmonic frequencies $\omega_p$ of the oscillatory source $\omega_{0}$ above the frequency given by the mass threshold $\omega_p \sim m_H \sim \sqrt{\lambda}$. This imposes a cutoff frequency, such that massive radiation is suppressed with increasing mass. The presence of these harmonics (with reducing amplitude) indicates that the radiation generation mechanism in the cases investigated is perturbative. 

\begin{figure*}
    \centering
    \includegraphics[width=\linewidth]{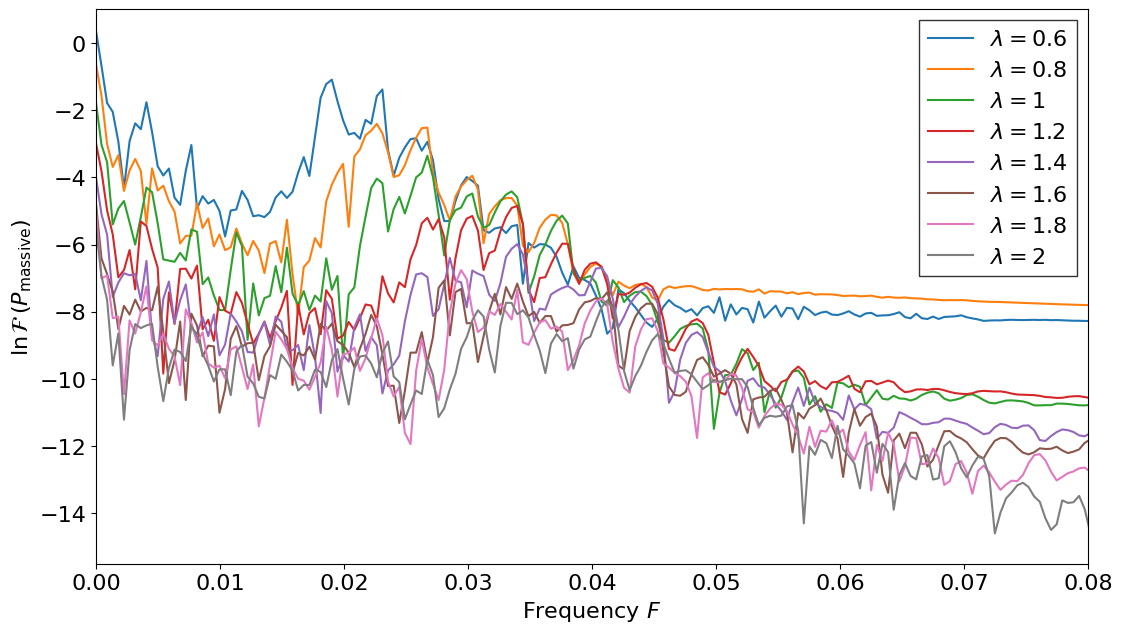}
    \includegraphics[width=\linewidth]{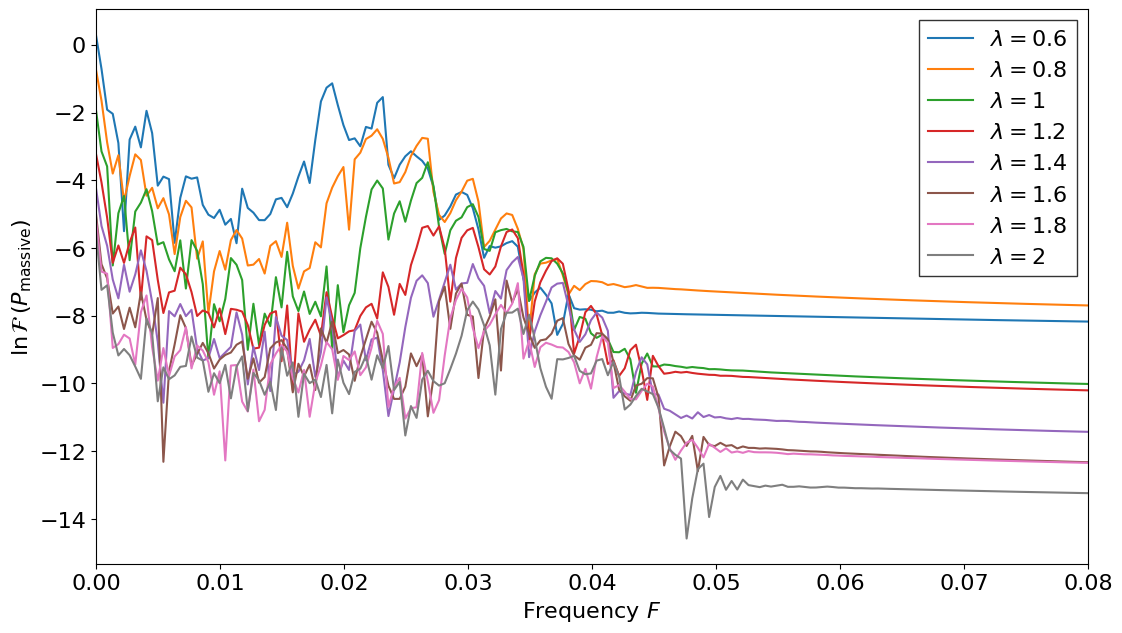}
    \caption{Spectrum of the massive radiation $P_{\mathrm{massive}}$ emitted from a range of $\lambda$ for $A_{\,\rm{rel}}=0.5$. The top spectrum is obtained using fixed grid simulations and no Kreiss-Oliger damping, and the bottom using AMR with $\sigma=1$. We observe a clear exponential decay in the spectrum, particularly in the fixed grid case. In the AMR simulations, higher modes tail off slightly, due either to radiation trapping by the refinement levels or the use of Kreiss-Oliger damping. Modes up to $p\sim 16$ can be distinguished in the fixed grid simulation, with a maximum $p \sim 12$ for the AMR simulation. The lowest plotted $\lambda=0.6$ emits primarily in the $p=5$ mode.}
    \label{spectrum}
\end{figure*}

In order to model the $\lambda$-dependence of the massive radiation from our sinusoidal string configurations, we must first examine the power spectrum of the radiation. This is given in Figure \ref{spectrum}, which shows the spectrum of $P_{\mathrm{massive}}$ for a range of $0.6 \leq \lambda \leq 2$. We observe two key features; first, we identify that the radiation is emitted in distinct harmonics, shown by the peaks in the spectrum evenly distributed in frequency $\sim \sqrt{\lambda}$. Second, we observe that, to leading order, the magnitude of the radiation falls off exponentially with $\sqrt{\lambda}$. This is perhaps unsurprising due to the perturbative nature of the source and expectations for massive radiation. We therefore introduce an exponential power of $\sqrt{\lambda}$ in any mass-dependent analytic model of the propagating radiation, for these and similar configurations.

To model the $\lambda$-dependence of our radiating string configurations, we build on the simple phenomenological model proposed in \cite{Olum2000}. This model describes radiation from sinusoidal Abelian-Higgs strings in terms of the local radius of curvature of the string $R$ at its maximum amplitude. This model and the corresponding simulations in \cite{Olum2000} consider an infinitely long string (i.e. with periodic boundary conditions in the $z$ direction) with a fixed relative amplitude $A=L/2$, for sinusoidal perturbations with varying $L$. It is postulated that an element $\mathrm d l$ of momentarily stationary curved string emits an element of energy $\mathrm{d}E \propto e^{-\alpha_C R}\,\mathrm{d}l$, where $\alpha_C$ is a constant.\footnote{This model for the Abelian-Higgs string is justified in \cite{Olum2000} using the exponential radial fall-off of the string profile. However, we observe from our results that it is also appropriate for global strings, whose profile has a $1/r^2$ fall-off. This is likely because radiated energy is not related to the string profile itself, but to the dynamics of the source. This concept has been studied in the context of gravitational wave source modelling, such as \cite{Sperhake_2017} and \cite{Rosca_Mead_2020}.} The energy radiated per period $E_{\rm rad}$ is then given by
\begin{equation}\label{exponential}
    E_{\rm rad}(L) \propto \int_{0}^L{e^{-\alpha_C R}\,\mathrm{d}l} \,.
\end{equation}
This can be simplified by considering only the string elements with the highest curvature, which contribute the most to the integral. For a sinusoidal curve in the regions of highest curvature,
\begin{equation}
    R \approx \frac{L^2}{A\,(2\pi)^2} + \left(\frac{6A\pi^2}{L^2} + \frac{1}{2A}\right)z^2\,,
\end{equation}
which for convenience we take to be around $z=0$. Substituting and performing the integral for $A=L/2$, we obtain
\begin{equation}
    E_{\rm rad}(L) \propto \sqrt{L}\exp{\left[-\alpha_C \frac{L}{2\pi^2}\right]}\,,
\end{equation}
where the approximation $dl \approx dz$ has been used. The factor of $R$ in the exponent captures the decrease in radiation suppression with increasing string curvature (decreasing radius of curvature). 

We adapt this model by incorporating a variable amplitude, using the relative amplitude $A_{\,\rm{rel}} = 4\,A / L$ (i.e. transforming $L\rightarrow 4L/A_{\,\rm{rel}}$), and introducing the mass dependence $m_H = \sqrt{\lambda}$ observed in Figure \ref{spectrum} into the exponent. We obtain the energy loss per period
\begin{equation}\label{newenergy}
    E(L,A_{\,\rm{rel}},\lambda) \propto f(L, A_{\,\rm{rel}})\exp{\left[-\gamma\sqrt{\lambda}\frac{L}{A_{\,\rm{rel}}}\right]}\,,
\end{equation}
where $f(L, A_{\,\rm{rel}})$ is a function of $L$ and $A_{\,\rm{rel}}$ and $\gamma$ is a constant to be determined. This can be compared directly with our measurements of $P_\mathrm{massive}$.

\begin{figure}[!]
    \centering
    \includegraphics[width=\linewidth]{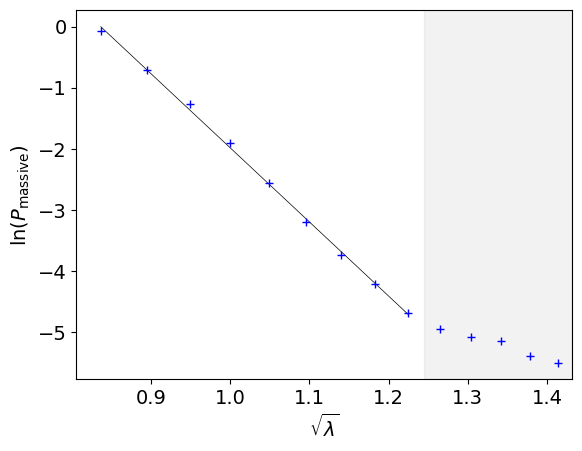}
    \caption{Massive radiation $P_{\rm massive}$ integrated over a diagnostic cylinder at $R=64$ from $90 \leq t\leq 228$ for $A_{\,\rm{rel}}=0.5$ and a range of $0.7 \leq \lambda \leq 2$. The black line indicates an exponential fit to the data for $0.7 \leq \lambda \leq 1.5$. The greyed-out area has not been included in the fit as it has been affected by numerical artefacts from the mesh refinement.}
    \label{L32_massive_exp}
\end{figure}

Figure \ref{L32_massive_exp} shows the massive radiation $P_{\mathrm{massive}}$ integrated over a diagnostic cylinder at $R=64$ over time from $90 \leq t \leq 228$ for $A_{\,\rm{rel}}=0.5$ and $L=32$. We first observe that, for $\lambda \gtrsim 1.5$, numerical effects have clearly become dominant, overtaking the physical energy loss.\footnote{We note that these numerical effects at $\lambda \gtrsim 1.5$ are not significant when compared to the massless radiation, as the relative magnitude is small; for example, as discussed in Section \ref{relenergy}, for $A_{\,\rm{rel}}=0.5$ and $\lambda \approx 1$, the massive radiation is over $1000\times$ smaller than massless radiation. This means these numerical artefacts should not have a significant effect on the string evolution overall.} We therefore exclude these points from our analysis. We also exclude points with $\lambda \lesssim 0.6$, as we observe from the evolution of the string amplitude that, at this point, internal mode oscillations of the string begin to interfere with the macroscopic oscillation of the string, such that the amplitude of oscillation no longer decays faster with decreasing $\lambda$ (i.e. lower than expected radiative decay). Finally and importantly, we note that, for the range of $\lambda$ analysed here, the measured $\sim 15\%$ decrease in $P_{\mathrm{massive}}$ measured for similar $\lambda$ by AMR simulations compared to fixed grid simulations, although slightly affecting the magnitude of the radiation, does not affect the gradient of the plot.

\begin{figure}[t]
    \centering
    \includegraphics[width=\linewidth]{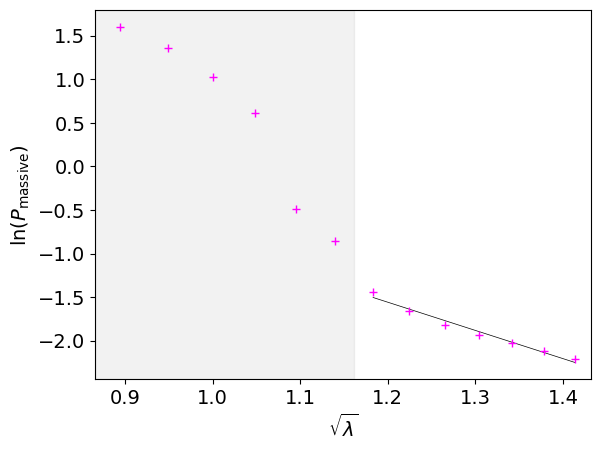}
    \caption{Massive radiation $P_{\mathrm{massive}}$ integrated over a diagnostic cylinder at $R=64$ from $90 \leq t\leq 360$ for $A_{\,\rm{rel}}=0.875$ and a range of $0.8 \leq \lambda \leq 2$. The black line indicates an exponential fit to the data for $1.6 \leq \lambda \leq 2$. The greyed-out area has not been included in the fit as it is significantly affected by nonlinear excitations, including internal mode oscillations and large amplitude higher harmonics.}
    \label{L16_massive_summary}
\end{figure}

From Figure \ref{L32_massive_exp}, we observe a clear exponential decay with $\sqrt{\lambda}$ from $0.7 \lesssim \lambda \lesssim 1.5$, consistent with (\ref{newenergy}). We perform a least-squares regression, also plotted in Figure \ref{L32_massive_exp}, to obtain $\gamma$ from the gradient,
\begin{equation}
    \gamma = 0.190 \;(\pm\; 0.003)\,.
\end{equation}
The root mean square error, quoted in brackets, is calculated from the accuracy of the fit and does not take into account the choice of points or other numerical uncertainties. We also note that an exponential model with a $\lambda$ dependence also provides a good (although less good) fit to the data, giving a root mean square error of 2.0\% relative to the measured gradient, compared to 1.4\% for the $\sqrt{\lambda}$ model. However, this $\lambda$-dependence would require a dependence on the radius of curvature of $R^2$ in the exponent to be dimensionally consistent.

%This fitting gives a root mean square error of 2.0\% relative to the measured gradient, compared to 1.4\% for the $\sqrt{\lambda}$ model. The graph for this fit is provided by Figure \ref{L32_massive_exp-appndx} in Appendix \ref{AppendixB}.

Figure \ref{L16_massive_summary} shows the extracted massive signal $P_{\mathrm{massive}}$ for the configuration $A_{\,\rm{rel}}=0.875$ and $L=16$. In this case, we integrate from $90 \leq t \leq 360$, as the radiated signal is longer, so a longer integration is required to provide an accurate picture of the decay. In this non-linear regime, finite width effects affect the evolution up to $\lambda \lesssim 1.3$, higher than for $A_{\,\rm{rel}}=0.5$, so these points are excluded. However, the higher overall magnitude of the radiation means that higher $\lambda$ can be investigated without being affected by numerical artefacts. We again calculate the gradient of the decay using a least-squares best fit to an exponential model, obtaining
\begin{equation}
%    \gamma = 0.229\; (\pm\; 0.026)\,.
%    \gamma = 0.145\; (\pm\; 0.002)\,.
    \gamma = 0.177\; (\pm\; 0.012)\,,
\end{equation}
with a relative error of $7\%$. The parameter $\gamma$ is consistent between the two datasets within two standard errors. This offers further evidence that an exponential decay model is consistent between these two rescaled string configurations. 

We also note that both of these $\gamma$ values are consistent with the investigation into energy decay from Abelian-Higgs strings presented in \cite{Olum2000} (within three standard deviations for $A_{\,\rm{rel}}=0.5$ and one standard deviation for $A_{\,\rm{rel}}=0.875$). For their setup, these authors obtain an equivalent of $\gamma=0.183$. A priori, we may not expect the decay constant for global and Abelian-Higgs strings to agree, as local strings have both scalar and gauge channels available for massive radiative decay at low $\lambda$, while the global strings have backreaction from massless radiation. This may suggest that of these two, scalar massive radiation could be the dominant radiative channel for their specific local string configuration, though would require further investigation.

% and the intercept $\ln{(K\sqrt{LA_{\,\rm{rel}}})} = 9.80 \;(\pm\; 0.41)$, where $K$ is a multiplicative constant \ad{do we use multiplicative constant from (45)?}. 

%\ad{*update} In the above analysis, the first two points have been excluded due to finite width effects that cause slower radiative decay. As discussed in \cite{Drew2019}, suppression at low $\lambda$ is likely due to massive internal oscillations within the string core interfering with the radiation mechanism. 

 %\ad{Fundamental frequency depends on the amplitude - look into this if time. Comparing with the massless paper, omega = 2pi/alpha L seems to become more accurate as lambda increases, otherwise there is damping through energy loss.}

\begin{figure}[!]
    \centering
    \includegraphics[width=\linewidth]{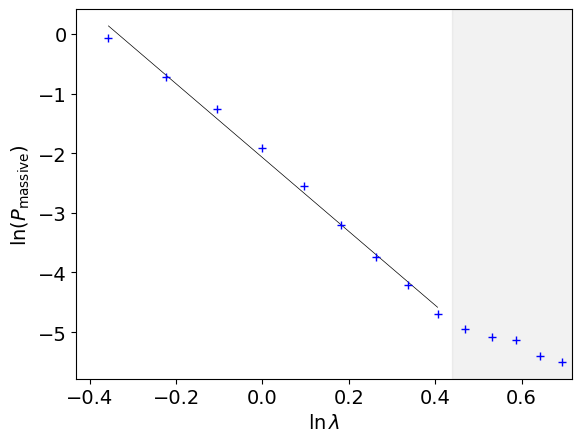}
    \caption{Massive radiation $P_{\rm massive}$ integrated over a diagnostic cylinder at $R=64$ from $90 \leq t\leq 228$ for $A_{\,\rm{rel}}=0.5$ and a range of $0.7 \leq \lambda \leq 2$. The black line indicates a power law fit to the data for $0.7 \leq \lambda \leq 1.5$. The greyed-out area has not been included in the fit as it has been affected by numerical artefacts from the mesh refinement.}
    \label{L32_massive_summary}
\end{figure}

Some contrasting alternative models, such as that proposed in \cite{Vincent:1997cx}, predict a power law decay of string radiation via a primary radiation channel of massive particles. To provide a comparison, we therefore fit a power law
\begin{equation}\label{powerlaw}
    E(L,A_{\,\rm{rel}},\lambda) \propto g(L, A_{\,\rm{rel}})(\sqrt{\lambda})^{\,-\gamma_{\rm{pow}}}
\end{equation}
to the same data. The fitting for $A_{\,\rm{rel}}=0.5$ and $L=32$ is given in Figure \ref{L32_massive_summary}, from which we obtain the decay coefficient
\begin{equation}
    \gamma_{\mathrm{pow}} = 6.2\; (\pm 0.2)\,.
\end{equation}
We note that the root mean square error is $2.9\%$  relative to the measured gradient, demonstrating a somewhat less consistent fit than the previous exponential decay model (\ref{exponential}). Nevertheless, this large power law suppression means that, for this configuration, particle radiation will not be a significant decay channel when extrapolating to high $\lambda$.

The power law fit at higher amplitude $A_{\,\rm{rel}}=0.875$ and $L=16$ yields the decay coefficient 
\begin{equation}
    \gamma_{\mathrm{pow}} = 2.1\; (\pm 0.1)\,,
\end{equation}
with a relative error of $6\%$. In contrast to the exponential model (\ref{exponential}) which provides a consistent exponent across the two regimes, here we identify a different power law in this more nonlinear regime. It may be possible to introduce further modelling to improve the fit of the power law model (\ref{powerlaw}), but the simple exponential model (\ref{exponential}) remains physically well-motivated and consistent with prior expectations about massive radiation. A power law suppression model to describe the massive radiation decay is not excluded by this work and may be appropriate in some regimes (e.g. for the extreme case $A_\mathrm{rel} \sim \varepsilon_\mathrm{eff} > 1$).

%and the intercept $\ln{(K\sqrt{LA_{\,\rm{rel}}})} = 7.64 \; (\pm\; 0.62)$. 

%%%We expect the ratio between the intercepts $r$ to be given by $\ln{(32\times0.5)} / \ln{(16\times0.875)} \approx 1.1$. We obtain $r = 1.28 \pm 0.18$, which again agrees with the prediction to within one standard error. \ad{There is an error in the intercept analysis here, we shouldn't just be taking the ratios I don't think. More detailed prefactor improves fit, but not great agreement.}

%\begin{figure}[!]
%    \centering
%    \includegraphics[width=\linewidth]{L=16massive_loglog_thresholds_sqrt_fit_2_accurate_damping.png}
%    \caption{\ps{I would not bother with this figure!} Massive radiation $P_{\mathrm{massive}}$ integrated over a diagnostic cylinder at $R=64$ from $90 \leq t\leq 360$ for $A_{\,\rm{rel}}=0.875$ and a range of $0.8 \leq \lambda \leq 2$. The black line indicates a power law fit to the data for $1.4 \leq \lambda \leq 2$. The greyed-out area has not been included in the fit as it is significantly affected by internal mode oscillations.}
%    \label{L16_massive_summary_power_law}
%\end{figure}

%We therefore conclude that, despite the increased complexity of the higher harmonic radiation, massive radiation is indeed exponentially suppressed even for highly non-linear configurations, in regimes where the numerics have sufficient resolution to characterise it accurately. Further investigation is required to characterise the exponents and prefactors in (\ref{newenergy}) more precisely and to understand the nature of the decay for highly relativistic configurations with large amplitude $\varepsilon \approx \mathcal{O}(1)$. \ad{edit}

\section{Conclusion and Future Work}\label{conclusion}

We have presented an in-depth investigation of the massive radiation from sinusoidal configurations of global cosmic strings using the adaptive mesh refinement code, GRChombo. We have investigated strings with $0.3 \leq \lambda \leq 10$ primarily with two relative amplitudes, $A_{\,\rm{rel}}=0.5$ ($\varepsilon = 0.68$) and $A_{\,\rm{rel}}=0.875$ ($\varepsilon = 0.96$). 

We first presented convergence tests for strings with $\lambda=1$, $\lambda=2$ and $\lambda=10$, both using AMR and a fixed grid to facilitate a comparison. Both numerical approaches converge when the string core $\delta \sim 1/\sqrt{\lambda}$ is resolved by $\gtrsim 4$ grid points. The resolution required for convergence of a fixed grid simulation is therefore the same as the resolution required on the finest refinement level of the AMR simulations. For the configurations studied, we find that the total massive radiation emitted $P_{\mathrm{massive}}$ can sometimes be lower in AMR simulations than in the fixed grid case. This is likely due to trapping of some radiation by the refinement level boundaries, as well as implicit damping from the averaging scheme between grid levels. This conclusion is supported by the fact that the order of convergence decreases as $\lambda$ increases, indicating that numerical effects are becoming significant. The decomposition of modes is largely unaffected by this trapping, aside potentially from those at very high frequencies, which have a very small amplitude.

We have determined the Fourier decomposition of the massive modes, focusing primarily on two scenarios. First, we have examined the mode decomposition of $\lambda=1$ and $\lambda=10$ strings for different amplitudes. We have found that the massive radiation emitted from global strings is considerably more complex than the massless radiation, consisting of low magnitude, high frequency modes with comparable amplitude. We have found that for configurations up to $A_\mathrm{rel} \sim \varepsilon \sim 1$, massive radiation is significantly suppressed compared to the massless channel, making up at most $1\%$ of the total radiation, even for low $\lambda$. For extreme nonlinear amplitudes $A_\mathrm{rel} \sim \varepsilon_\mathrm{eff} \sim 1.5$ and $\lambda < 1$, it was possible to have massive radiation at a comparable magnitude to massless radiation (on the first oscillation).

Second, we have performed a finely-spaced scan of $\Delta \lambda=0.1$ to determine the $\lambda$-dependence of the massive radiation. We have observed for the configurations studied that massive radiation is emitted in distinct harmonics of the fundamental frequency of the string. This is predicted analytically by solving the massive Klein-Gordon equation, which also predicts the presence of a mass-dependent cutoff frequency $\omega_p = 2\pi p_\mathrm{min} / \alpha L$ below which massive modes cannot propagate. The lowest propagating harmonic is defined by $p _{\rm min} \gtrsim m_H/\Omega_z\, \approx \sqrt\lambda/\Omega_z$. We have confirmed the presence of this cutoff frequency by performing Fourier analyses of the massive spectrum for $0.3 \leq \lambda \leq 2$, for both $A_{\,\rm{rel}}=0.5$ and $A_{\,\rm{rel}}=0.875$, and have demonstrated the presence of several distinct harmonics in the power spectrum for $A_{\,\rm{rel}}=0.5$.

Finally, we have used this finely-spaced scan to estimate the $\lambda$-dependence of the massive radiation spectrum. For $A_\mathrm{rel} = 0.5$, we have demonstrated that the spectrum is characterised by an exponential falloff $P_\mathrm{massive} \propto e^{-\gamma\sqrt\lambda}$, and is not as well described by a simple power law model. In either case, this means that, for similar configurations, massive radiation will not be an important decay mechanism for oscillating strings. For $A_\mathrm{rel} = 0.875$ ($\varepsilon \sim 1$), we observe that the mass-dependence of the spectrum again fits well with the original exponential model (\ref{newenergy}) with the same falloff.  However, the power law model requires a different exponent $P_\mathrm{massive} \propto (\sqrt{\lambda})^{-2}$ in this regime.  Further investigation is required to determine whether or not a viable massive radiation channel is available in realistic cosmological networks. This will prove to be important for future numerical relativistic simulations of string evolution and predictions of their gravitational wave spectra, as well as estimates of the axion mass using simulations of axion string networks.

One of the key conclusions from this work is that, as long as careful consideration is given to the initial conditions and other numerical parameters, AMR is very useful for studying cosmic strings. As long as the string core is appropriately resolved, we are able to resolve even high harmonics of the massive propagating radiation. As the wavelength of the lowest propagating mode is offset by the particle mass, this should also be true for even higher $\lambda$ (until the resolution in time of the base grid becomes the limiting factor). The impact of numerical effects on the magnitude of the massive radiation is relatively small, especially when compared to the dominant massless radiation, and the potential causes of these are currently being addressed by the GRChombo collaboration. 

Further to this, we also observe a significant saving in computational time and resources when using AMR compared to a fixed grid, particularly for high $\lambda$. For example, for the $\lambda=10$ convergence test presented, the $\Delta x = 0.0625$ fixed grid simulation took a few days to run on 4096 CPUs, not including time spent in job queues, whereas the AMR run with $\Delta x_{l_{\rm max}} = 0.0625$ took a few hours on 128 CPUs. This means that accurate simulations of cosmic string networks with higher $\lambda$ than current fixed grid simulations, for example $\lambda=10$, can be performed up to $1000\times$ faster with AMR than using a fixed grid. We plan to exploit this capability in future work.

\hspace{1cm}

\section{Acknowledgements}

We are grateful for useful conversations with Michalis Agathos, Josu Aurrekoetxea, Katy Clough, José Ricardo Correia, Eugene Lim, Ulrich Sperhake and Weiqun Zhang. We would also like to thank and acknowledge the GRChombo team (http://www.grchombo.org/). We are particularly grateful to Kacper Kornet and Miren Radia for invaluable technical computing support.

We would like to acknowledge the support of the Intel Visualization team, led by Jim Jeffers, notably the collaboration on in-situ visualization with Carson Brownlee. 
We further acknowledge the CNRS-FTC Cooperation Grant IEA 2020 302614 and the organizers of the workshop `Cosmic Topological Defects: Dynamics and Multimessenger Signatures' at the Lorentz Center, Leiden, for facilitating helpful and enlightening discussions.

This work was undertaken on the COSMOS supercomputer at DAMTP, University of Cambridge, funded by BEIS National E-infrastructure capital grants ST/J005673/1 and STFC grants ST/H008586/1, ST/K00333X/1, the Fawcett supercomputer at DAMTP funded by STFC Consolidated Grant ST/P000673/1, and the Cambridge CSD3 part of the STFC DiRAC HPC Facility (www.dirac.ac.uk). The DiRAC component of CSD3 was funded by BEIS capital funding via STFC capital grants ST/P002307/1 and ST/R002452/1 and STFC operations grant ST/R00689X/1.

AD is supported by a Junior Research Fellowship (JRF) at Homerton College, University of Cambridge. Part of this work was undertaken whilst AD was supported by an EPSRC iCASE Studentship in partnership with Intel (EP/N509620/1, Voucher 16000206). EPS acknowledges funding from STFC Consolidated Grant ST/P000673/1.

\bibliography{Paper1new}

\begin{thebibliography}{10}

\bibitem{Kibble1976}
T.~W.~B. Kibble.
\newblock {Topology of Cosmic Domains and Strings}.
\newblock {\em J. Phys. A: Math. Gen}, 9(8):1387--1398, 1976.

\bibitem{Vilenkin:2000jqa}
A.~Vilenkin and E.~P.~S. Shellard.
\newblock {\em {Cosmic Strings and Other Topological Defects}}.
\newblock Cambridge University Press, 1994.

\bibitem{Peccei1977a}
R.~D. Peccei and H.~R. Quinn.
\newblock {CP conservation in the presence of pseudoparticles}.
\newblock {\em Phys. Rev. Lett.}, 38(25):1440--1443, 1977.

\bibitem{Davis1986}
R.~L. Davis.
\newblock Cosmic axions from cosmic strings.
\newblock {\em Phys. Lett. B}, 180(3):225--230, 1986.

\bibitem{Abbott_2021}
R.~Abbott et~al.
\newblock Constraints on cosmic strings using data from the third advanced
  {LIGO}{\textendash}virgo observing run.
\newblock {\em Phys. Rev. Lett.}, 126(24):241102, 2021.

\bibitem{Auclair_2020}
Pierre~Auclair et~al.
\newblock Probing the gravitational wave background from cosmic strings with
  {LISA}.
\newblock {\em Journal of Cosmology and Astroparticle Physics}, 2020(04):034,
  2020.

\bibitem{LISA}
Pierre Auclair~et al.
\newblock Cosmology with the laser interferometer space antenna.
\newblock arXiv:2204.05434, 2022.

\bibitem{NANOGRAV}
Z.~Arzoumanian et~al.
\newblock The {NANOGrav} 11 year data set: Pulsar-timing constraints on the
  stochastic gravitational-wave background.
\newblock {\em The Astrophysical Journal}, 859(1):47, 2018.

\bibitem{Olum2000}
K.~D. Olum and J.~J. Blanco-Pillado.
\newblock Radiation from cosmic string standing waves.
\newblock {\em Phys. Rev. Lett.}, 84(19):4288--4291, 2000.

\bibitem{Moore:1998}
J.~N. Moore and E.~P.~S. Shellard.
\newblock On the evolution of abelian-higgs string networks.
\newblock arXiv:hep-ph/9808336, 1998.

\bibitem{Vincent:1997cx}
G.~Vincent, N.~D. Antunes, and M.~Hindmarsh.
\newblock Numerical simulations of string networks in the abelian-higgs model.
\newblock {\em Phys. Rev. Lett.}, 80(11):2277--2280, 1998.

\bibitem{Gorghetto_2021}
Marco Gorghetto, Edward Hardy, and Giovanni Villadoro.
\newblock More axions from strings.
\newblock {\em {SciPost} Physics}, 10:050, 2021.

\bibitem{Hindmarsh_2021}
Mark Hindmarsh, Joanes Lizarraga, Asier Lopez-Eiguren, and Jon Urrestilla.
\newblock Comment on `more axions from strings'.
\newblock arXiv:2109.09679, 2021.

\bibitem{Buschmann_2022}
Malte~Buschmann et~al.
\newblock Dark matter from axion strings with adaptive mesh refinement.
\newblock {\em Nature Communications}, 13(1049), 2022.

\bibitem{Saurabh_2020}
Ayush Saurabh, Tanmay Vachaspati, and Levon Pogosian.
\newblock Decay of cosmic global string loops.
\newblock {\em Phys. Rev. D}, 101(8):083522, 2020.

\bibitem{Hindmarsh2021}
Mark~Hindmarsh et~al.
\newblock Approach to scaling in axion string networks.
\newblock {\em Phys. Rev. D}, 103(10):103534, 2021.

\bibitem{Blanco_Pillado_2018}
Jose~J. Blanco-Pillado, Ken~D. Olum, and Xavier Siemens.
\newblock New limits on cosmic strings from gravitational wave observation.
\newblock {\em Phys. Lett. B}, 778:392--396, 2018.

\bibitem{Matsunami_2019}
Daiju Matsunami, Levon Pogosian, Ayush Saurabh, and Tanmay Vachaspati.
\newblock Decay of cosmic string loops due to particle radiation.
\newblock {\em Phys. Rev. Lett.}, 122(20):201301, 2019.

\bibitem{Auclair_2020a}
Pierre Auclair, Dani{\`{e} }le~A. Steer, and Tanmay Vachaspati.
\newblock Particle emission and gravitational radiation from cosmic strings:
  Observational constraints.
\newblock {\em Phys. Rev. D}, 101(8):083511, 2020.

\bibitem{Hindmarsh_2020}
Mark Hindmarsh, Joanes Lizarraga, Asier Lopez-Eiguren, and Jon Urrestilla.
\newblock Scaling density of axion strings.
\newblock {\em Phys. Rev. Lett.}, 124(2):021301, 2020.

\bibitem{Auclair_2021}
Pierre Auclair, Konstantin Leyde, and Danièle~A. Steer.
\newblock A window for cosmic strings.
\newblock arXiv:2112.11093, 2021.

\bibitem{Hindmarsh_2021a}
Mark~Hindmarsh et~al.
\newblock Loop decay in abelian-higgs string networks.
\newblock {\em Phy. Rev. D}, 104(4):043519, 2021.

\bibitem{Hindmarsh_2022}
Mark Hindmarsh and Jun'ya Kume.
\newblock Multi-messenger constraints on abelian-higgs cosmic string networks.
\newblock arXiv:2210.06178, 2022.

\bibitem{Helfer2019}
T.~Helfer, J.~C. Aurrekoetxea, and E.~A. Lim.
\newblock Cosmic string loop collapse in full general relativity.
\newblock {\em Phys. Rev. D}, 99(10):104028, 2019.

\bibitem{Aurrekoetxea_2020}
Josu~C Aurrekoetxea, Thomas Helfer, and Eugene~A Lim.
\newblock Coherent gravitational waveforms and memory from cosmic string loops.
\newblock {\em Classical and Quantum Gravity}, 37(20):204001, 2020.

\bibitem{Aurrekoetxea_2022}
Josu~C. Aurrekoetxea, Pedro~G. Ferreira, Katy Clough, Eugene~A. Lim, and
  Oliver~J. Tattersall.
\newblock Where is the ringdown: Reconstructing quasinormal modes from
  dispersive waves.
\newblock {\em Phys. Rev. D}, 106(10):104002, 2022.

\bibitem{Drew2019}
A.~Drew and E.~P.~S. Shellard.
\newblock Radiation from global topological strings using adaptive mesh
  refinement: Methodology and massless modes.
\newblock {\em Phys. Rev. D}, 105(6):063517, 2022.

\bibitem{Andrade:2021}
Tomas Andrade~et al.
\newblock {{GRChombo}}: {{An}} adaptable numerical relativity code for
  fundamental physics.
\newblock {\em Journal of Open Source Software}, 6(68):3703, 2021.

\bibitem{Radia2021}
Miren~Radia et~al.
\newblock Lessons for adaptive mesh refinement in numerical relativity.
\newblock {\em Classical and Quantum Gravity}, 39(13):135006, 2021.

\bibitem{Kreiss:1973}
H.~Kreiss and J.~Oliger.
\newblock {\em Methods for the Approximate Solution of Time Dependant
  Problems}.
\newblock Number~10 in {{GARP Publication Series}}. {WMO}, {Geneva}, 1973.

\bibitem{Battye1993}
R.~A. Battye and E.~P.~S. Shellard.
\newblock Global string radiation.
\newblock {\em Nucl. Phys. B}, 423(1):260--304, 1994.

\bibitem{Battye1995}
R.~A. Battye and E.~P.~S. Shellard.
\newblock Radiative back reaction on global strings.
\newblock {\em Phys. Rev. D}, 53(4):1811--1826, 1996.

\bibitem{Sperhake_2017}
Ulrich~Sperhake et~al.
\newblock Long-lived inverse chirp signals from core-collapse in massive
  scalar-tensor gravity.
\newblock {\em Phys. Rev. Lett.}, 119(20):201103, 2017.

\bibitem{Rosca_Mead_2020}
Roxana Rosca-Mead et~al.
\newblock Core collapse in massive scalar-tensor gravity.
\newblock {\em Phys. Rev. D}, 102(4):044010, 2020.

\end{thebibliography}

%\clearpage

\appendix
\section{$\lambda=2$ and $\lambda=10$ Convergence Test Plots}\label{AppendixA}

\begin{figure}
    \centering
    \includegraphics[width=\linewidth]{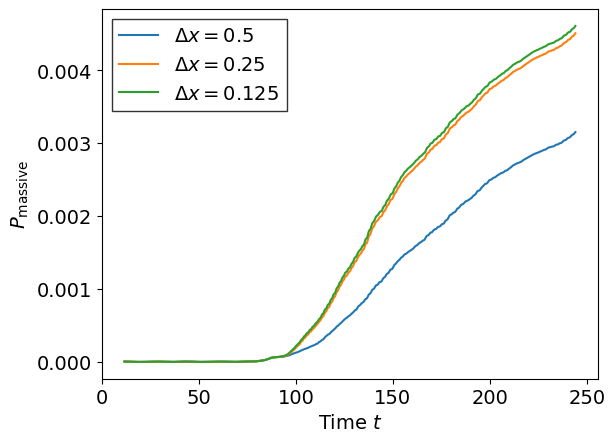}
    \includegraphics[width=\linewidth]{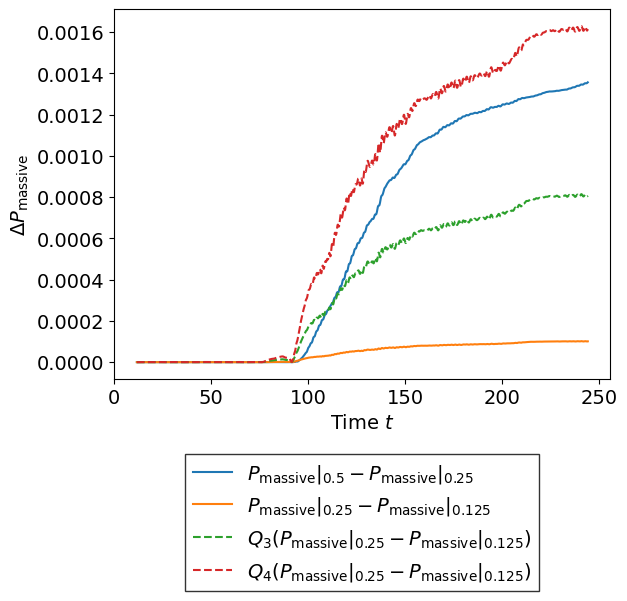}
    \caption{Absolute value (top) and convergence (bottom) of the energy emitted by massive radiation $P_{\mathrm{massive}}$ from a $\lambda=2$ string with initial amplitude $A_0=4$, measured on a cylinder at $R=64$ on a fixed grid for different refinements $\Delta x$ (test \textit{FG} in Table \ref{convergence_params}). The convergence plot shows the difference in the magnitude of $P_{\mathrm{massive}}$ between different resolutions, with the higher resolution results also plotted rescaled according to third- and fourth-order convergence.}
    \label{PoyntingFixedLambda2}
\end{figure}

\begin{figure}
    \centering
    \includegraphics[width=\linewidth]{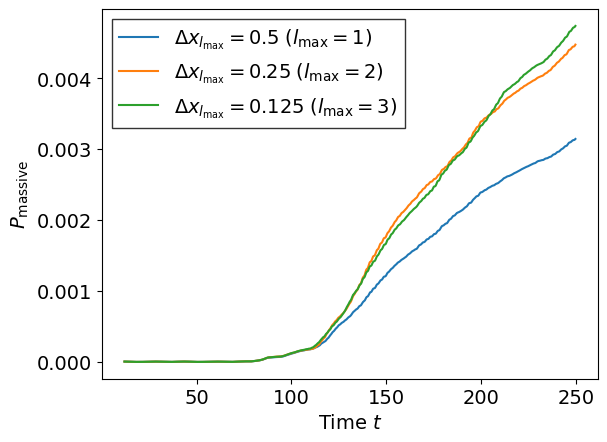}      
    \includegraphics[width=\linewidth]{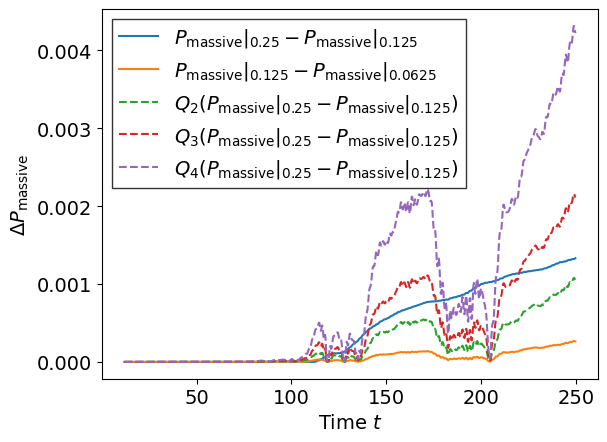}          
%    \includegraphics[width=0.5\textwidth]{lambda_2_max_2_to_5_energy_total_regrid_0.1.png}
%    \includegraphics[width=0.5\textwidth]{lambda_2_max_2_to_5_energy_convergence_regrid_0.1.png}
%    \caption{\textcolor{ForestGreen}{Absolute value (top) and and convergence (bottom) of the energy emitted by massive radiation $P_{\mathrm{massive}}$ from a $\lambda=2$ string with initial amplitude $A_0=4$, measured on a cylinder at $R=64$ using adaptive mesh refinement (test \textit{AMR} in Table \ref{convergence_params}). The grid resolutions on the finest refinement levels are $\Delta x_{l_{\rm max}}=0.25,\,0.125\,,0.0625\; \rm{and}\; 0.03125$. The convergence plot shows the difference in the magnitude of $P_{\mathrm{massive}}$ between different resolutions, with the higher resolution results also plotted rescaled according to sixth- and seventh-order convergence. The dashed black line in the top plot is drawn at $P_{\mathrm{massive}} = 0.005$, approximately the maximum $P_{\mathrm{massive}}$ measured for an identical setup but using a fixed grid in Figure \ref{PoyntingFixed}.}}
    \caption{Absolute value of the energy emitted by massive radiation $P_{\mathrm{massive}}$ from a $\lambda=2$ string with initial amplitude $A_0=4$, measured on a cylinder at $R=64$ using adaptive mesh refinement (test \textit{AMR} in Table \ref{convergence_params}). The convergence plot shows the difference in the magnitude of $P_{\mathrm{massive}}$ between different resolutions, with the higher resolution results also plotted rescaled according to second-, third- and fourth-order convergence.}
    \label{PoyntingAMRLambda2}
\end{figure}

\begin{figure}
    \centering
    \includegraphics[width=0.5\textwidth]{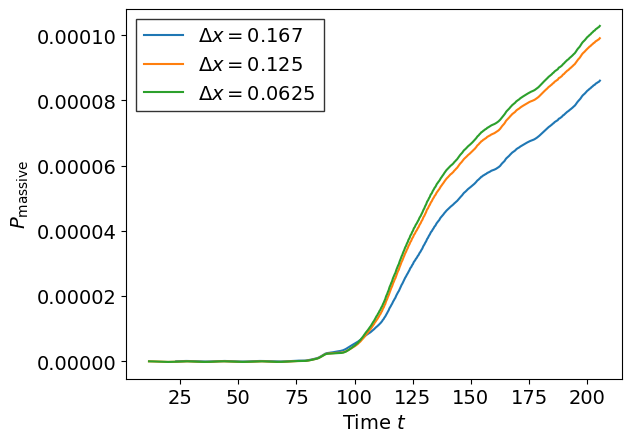}
    \includegraphics[width=0.5\textwidth]{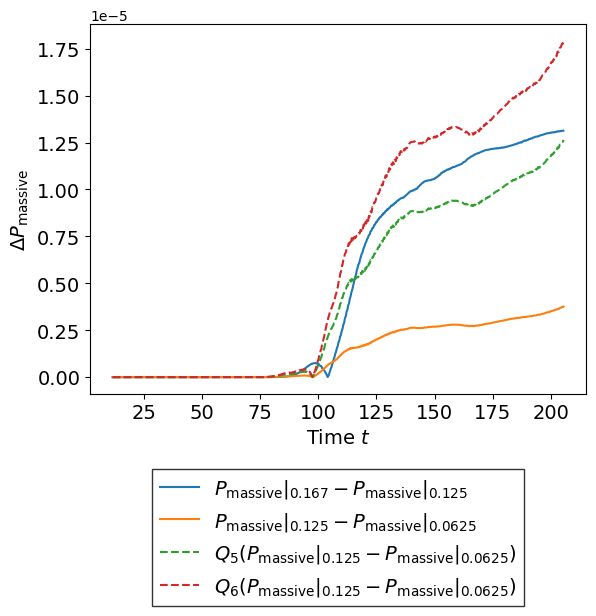}
    \caption{Absolute value (top) and convergence (bottom) of the energy emitted by massive radiation $P_{\mathrm{massive}}$ from a $\lambda=10$ string with initial amplitude $A_0=4$, measured on a cylinder at $R=64$ on a fixed grid for different refinements $\Delta x$ (test \textit{FG} in Table \ref{convergence_params}). The convergence plot shows the difference in the magnitude of $P_{\mathrm{massive}}$ between different resolutions, with the higher resolution results also plotted rescaled according to fifth- and sixth-order convergence.}
    \label{PoyntingFixedLambda10}
\end{figure}

\begin{figure}
    \centering
    \includegraphics[width=0.5\textwidth]{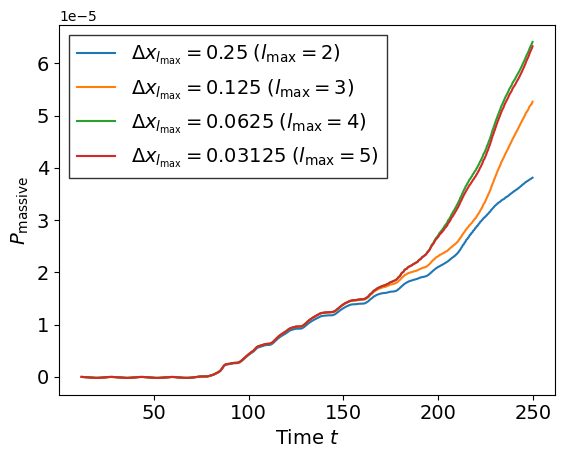}
    \includegraphics[width=0.5\textwidth]{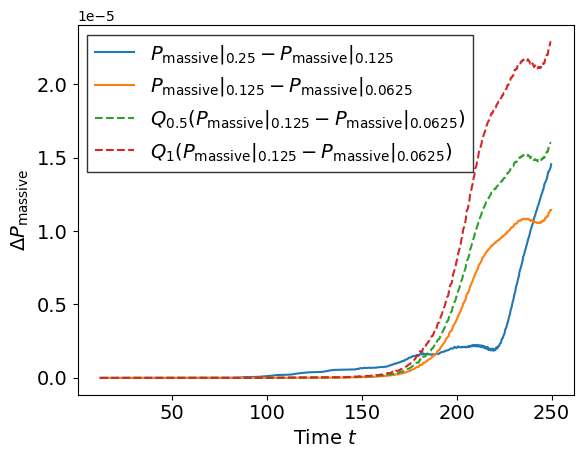}
    \caption{Absolute value of the energy emitted by massive radiation $P_{\mathrm{massive}}$ from a $\lambda=10$ string with initial amplitude $A_0=4$, measured on a cylinder at $R=64$ using adaptive mesh refinement (test \textit{AMR} in Table \ref{convergence_params}). The convergence plot shows the difference in the magnitude of $P_{\mathrm{massive}}$ between different resolutions, with the higher resolution results also plotted rescaled according to 0.5th- and first-order convergence.}
    \label{PoyntingAMRLambda10}
\end{figure}

Here we present the convergence test plots for $\lambda=2$ and $\lambda=10$ used in the analysis of Section \ref{convergencetestsresults}. Figures \ref{PoyntingFixedLambda2} and \ref{PoyntingAMRLambda2} show the $\lambda=2$ tests for a fixed grid and AMR respectively. Figures \ref{PoyntingFixedLambda10} and \ref{PoyntingAMRLambda10} show the fixed grid and AMR tests for $\lambda=10$.

\hspace{2cm}

\end{document}